\documentclass[10pt,pre,a4paper,twocolumn,superscriptaddress,showpacs]{revtex4-1}
\usepackage[latin1]{inputenc}
\usepackage[colorlinks=true]{hyperref}
\usepackage{amsmath,amssymb}
\usepackage{amsfonts}
\usepackage{graphicx}
\usepackage{color}
\usepackage[dvipsnames]{xcolor}

\begin{document}

\title{Athermal rheology of weakly attractive soft particles}

\author{Ehsan Irani} \affiliation{Institute for Theoretical Physics, Georg-August University of G\"ottingen, Friedrich-Hund Platz 1, 37077 G\"ottingen, Germany}
\author{Pinaki Chaudhuri} \affiliation{Institute of Mathematical
  Sciences, Taramani, Chennai 600 113, Tamil Nadu, India}
\author{Claus Heussinger} \affiliation{Institute for Theoretical
  Physics, Georg-August University of G\"ottingen, Friedrich-Hund
  Platz 1, 37077 G\"ottingen, Germany}

\begin{abstract}
We study the rheology of a soft particulate system where the
inter-particle interactions are weakly attractive. Using extensive
molecular dynamics simulations, we scan across a wide range of
packing fractions ($\phi$), attraction strengths ($u$) and imposed
shear-rates ($\dot{\gamma}$). In striking contrast to repulsive
systems, we find that at small shear-rates generically a fragile
isostatic solid is formed even if we go to $\phi \ll \phi_J$.
Further, with increasing shear-rates, even at these low $\phi$,
non-monotonic flow curves occur which lead to the formation of
persistent shear-bands in large enough systems. By tuning the damping
parameter, we also show that inertia plays an important role in
this process. Furthermore, we observe enhanced particle dynamics
in the attraction-dominated regime as well as a pronounced anisotropy
of velocity and diffusion constant, which we take as precursors to
the formation of shear bands.  At low enough $\phi$, we also observe
structural changes via the interplay of low shear-rates and attraction
with the formation of micro-clusters and voids.  Finally, we
characterize the properties of the emergent shear bands and thereby,
we find surprisingly small mobility of these bands, leading to
prohibitely long time-scales and extensive history effects in ramping
experiments.
  \end{abstract}

\maketitle
\section{Introduction}

Soft jammed materials (e.g. foams, grains, gels, emulsions, colloids,
etc.) are commonplace in nature and our daily lives, exhibiting a wide
range of rheological behaviour. Due to large scale industrial
applications and also abundance in natural phenomena, understanding
the flow properties of these soft materials is an area of intense
research. Despite the wide variety of materials, many such systems
exhibit the phenomenon of {\it jamming}, which is a non-equilibrium
transition \cite{vanHeckeJPCM2010, LiuNagelAnnRevCondMattPhys2010}
whereby the material becomes solid when the volume fraction, $\phi$,
of the constituent particles crosses a threshold, i.e. the jamming
point $\phi_J$.  Above $\phi_J$, a stress threshold (known as {yield
  stress}), needs to be exceeded to obtain a steadily flowing
state. In the context of rheology, the development of a finite yield
stress constitutes the onset of jamming at $\phi=\phi_J$
\cite{ClausPinakiSoftmatter2010}. To understand and develop theories for this phenomenon, systems with simplified particle interactions have served as models. In particular, interactions based on repulsive, and frictionless contacts have been studied extensively~\cite{vanHeckeJPCM2010,LiuNagelAnnRevCondMattPhys2010}. It is now well known that in granular systems, the presence of frictional interactions modify the jamming phase diagram~\cite{GrobPRE2014} and jamming becomes possible in a range of volume fractions depending on the preparation as well as on the inter-particle friction coefficient~\cite{daCruzPRE2005}. The role of attractive interactions in determining the rheology of such complex fluids has, however, only recently commenced~\cite{IraniPRL2014,PinakiPRE2012,SundaresanPRE2014,RahbariPRE2010,SinghPRE2014}.
\begin{figure}[h!tb]
\includegraphics[width=0.5\textwidth]{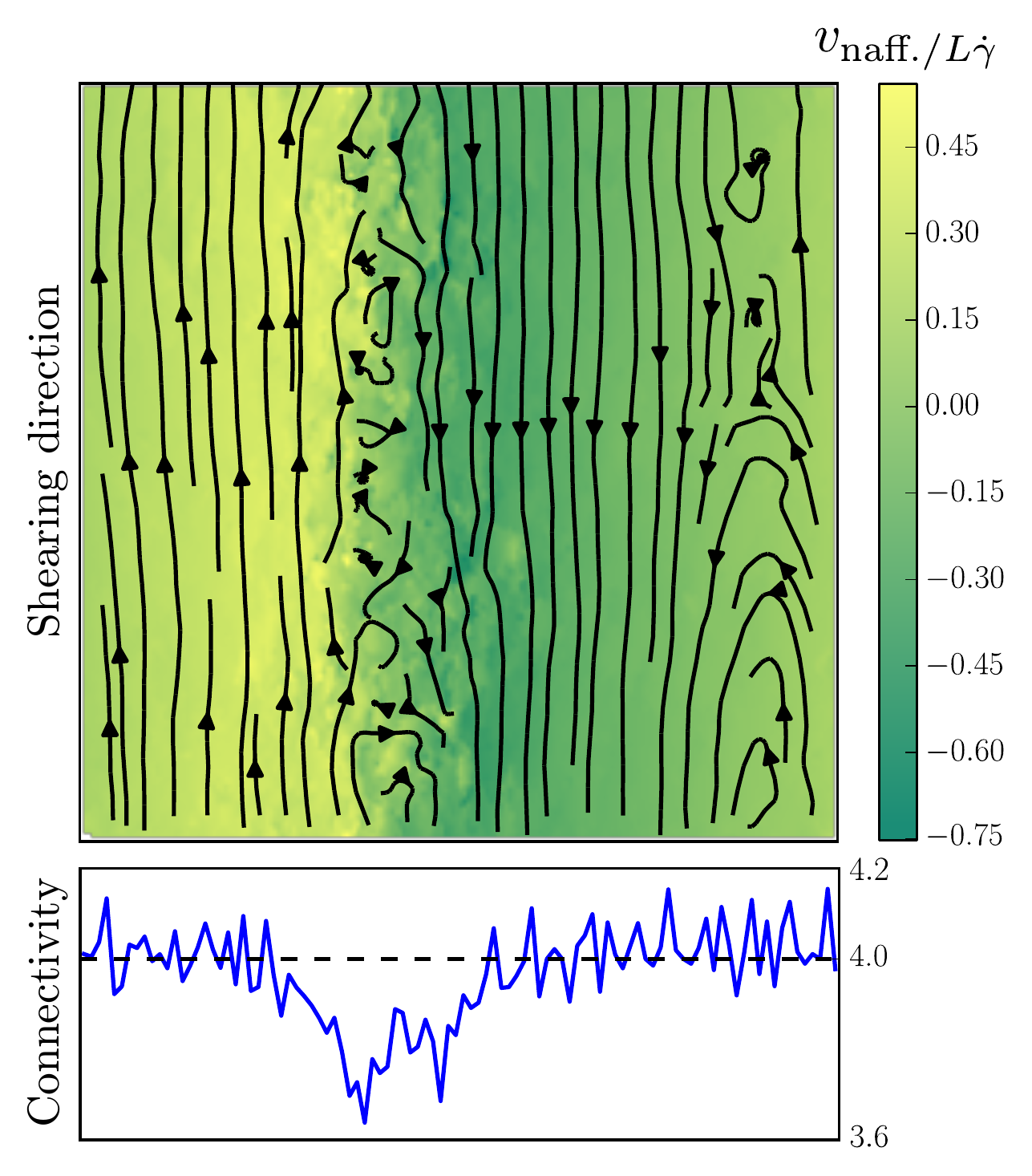}
	\caption{(Color online, top), A {snapshot of the non-affine velocity
            field} in a system exhibiting shear-banding, using
          $N=2\times{10^4}$ particles, $\phi=0.82$, $u=2\times10^{-5}$ and
          $\dot\gamma=2.5\times{10^{-6}}$. The coloring is based on the
          velocity in the shearing direction. Stream lines show the
          non-affine flow field. (Bottom), the corresponding
          connectivity profile. It reveals that in the solid band,
          connectivity fluctuates around the isostactic value but
          decreases in the fludized band.}
        \label{fig:snap-velf-z}
\end{figure}

In the case of granular materials, attraction can appear in different
ways, like the development of capillary bridges
\cite{HornbakerNature1997, HerminghausAdvPhys2005, MitaraiAdvPhys2006}
or van der Waals forces \cite{CastellanosAdvPhys2005,
  RoyerNature2009}, and it might change the rheology of the system
significantly. For example, a finite yield stress is observed in some
attractive systems at packing fractions much less than the repulsive
$\phi_J$ \cite{CoussotSoftMatter2007, Moller5139, GreggOHernPRL2008,
  RahbariPRE2010}. Furthermore, shear-banding, i.e. the occurrence of
spatially inhomogenoeous flow \cite{SchallShearBand2010,
  CoussotShearBand2009}, is often observed in these materials. While
shear localization has been reported in both repulsive and attractive
systems~\cite{Moller5139, BecuPRL2006}, it has now been established
that dense systems of soft repulsive particles, studied over
sufficiently long times, do not exhibit shear bands as a permanent
feature \cite{OvarlezEPL2010, DivouxPRL2010, BecuPRL2006}. For the
occurrence of permanent shear bands, non-monotonic constitutive laws
are necessary \cite{SchallShearBand2010, FieldingRepProgPhys2014},
which is not observed in repulsive systems. On the other hand, in an
earlier work \cite{IraniPRL2014}, we have demonstrated how such
non-monotonic flow curves can be obtained by including weak
attractive interactions. Therefore studying such systems can shed
light on physical mechanisms behind the observation of permanent flow
heterogeneities in athermal jammed materials.

In our previous work \cite{IraniPRL2014}, using numerical simulations,
we investigated the effect of weak attractive interactions on the
rheology of granular systems.  We observed the development of a
finite yield stress below the jamming point and also the occurrence
of non-monotonic flow curves leading to permanent shear bands (Fig.
\ref{fig:snap-velf-z}). We also proposed a simple theoretical model
to rationalise the observations, based on the competition between
shear-induced fluidization and the tendency of aggregation. In the
current study, we concentrate on extending the jamming phase diagram,
and demarcate the regimes where shear-banding can be observed. We
also discuss the associated flow properties, local structure and
particle dynamics at microscopic level in greater details, specifically
in the crossover from attraction-dominated to repulsion-dominated
regime.  Further, we investigate the properties of shear bands: the
behaviour of the interface as a function of external strain rate,
the dynamics of the entire band etc. These help in providing a more
coherent picture regarding the occurrence of shear-bands in attractive
systems.

The paper is organized as follows. First, we introduce the model
and outline the simulation method. This is done in Sect.~\ref{sec:model}.
Then, we discuss our results in details in Sect.~\ref{sec:results}.
This is carefully structured in order to sequentially discuss our
findings starting with analysis concerning macroscopic (system-level)
behavior (Sect.~\ref{sec:macroscopic-rheology}) and then connecting
to microscopic (particle level) aspects, separating out structural
(Sect.~\ref{sec:structure-factor}) and dynamical
(Sect.~\ref{sec:microscopic-dynamics}) observations.  We end our
discussions in Sect.~\ref{sec:char-shear-bands} with a systematic
analysis of the formation of flow heterogeneities.  At the end
(Sect.~\ref{sec:conclusion}), we conclude and discuss our results.

%

\section{Model}\label{sec:model}

\begin{figure}[h]
	\centering
	\includegraphics[width=0.45\textwidth]{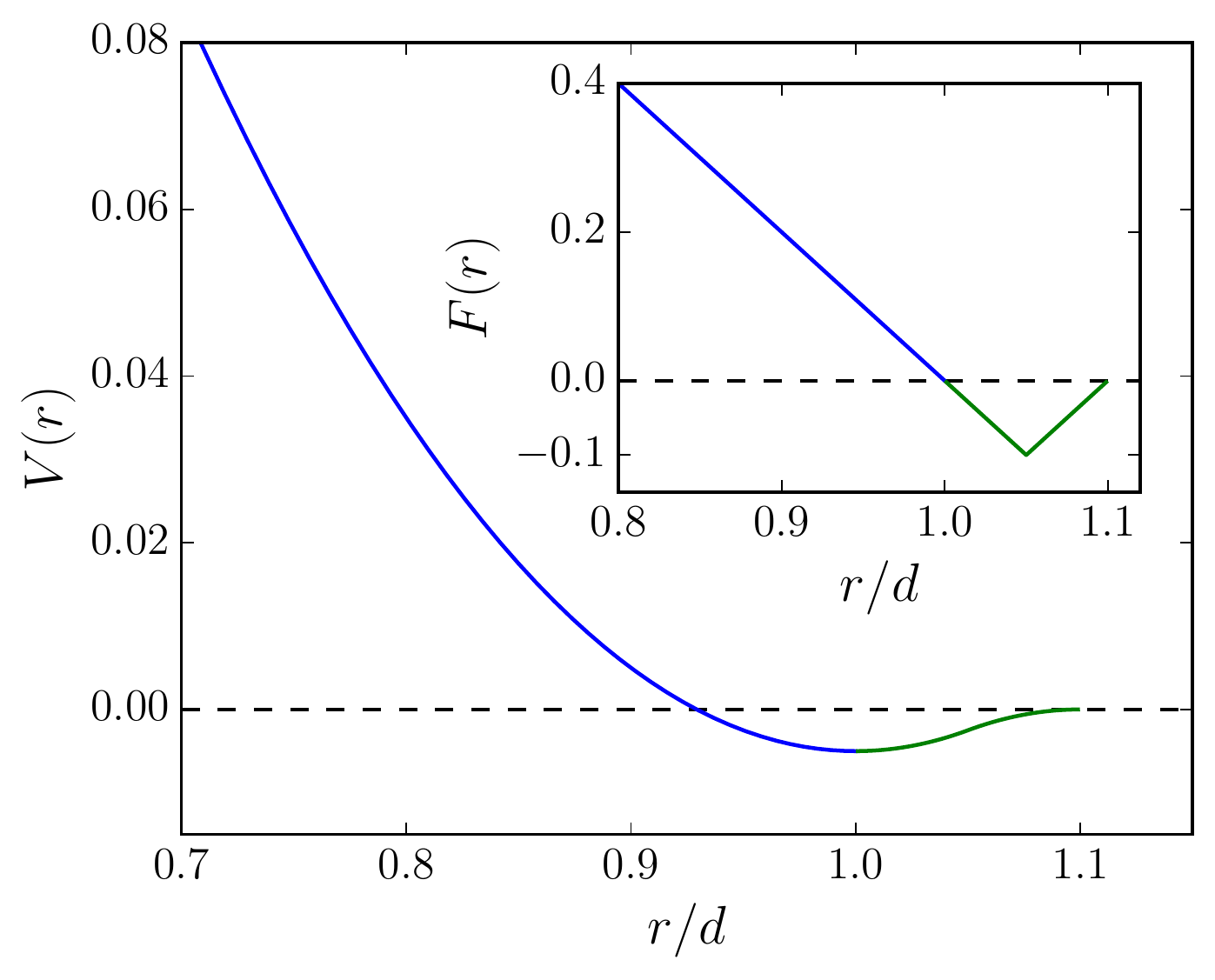}
	\caption{Interparticle potential $V(r)$, for $\epsilon=1$ and
          $u=0.05$. The inset is the corresponding elastic force. The
          attractive part is shown in green ($r/d > 1.0$).}
	\label{fig:epot-f}
\end{figure}

We consider a two-dimensional system of $N$  soft disks interacting via the following potential:
\begin{equation}
\label{eq:pot}
V(r_{ij})=
\begin{cases}
\epsilon\left[(1-\frac{r_{ij}}{d_{ij}})^2-2u^2\right],
& \frac{r_{ij}}{d_{ij}} < 1+u \\
-\epsilon\left[1+2u-\frac{r_{ij}}{d_{ij}}\right]^2,
& 1+u < \frac{r_{ij}}{d_{ij}} < 1+2u \\
0, & \frac{r_{ij}}{d_{ij}} > 1+2u \\
\end{cases}
\end{equation}
where $r_{ij}$ is the distance between the $i$th and $j$th particles,
and $d_{ij}=(d_i+d_j)/2$ is the summation of their radii. Thus, there
exists a harmonic repulsive interaction when the particles overlap,
$r_{ij}<d_{ij}$. Additionally, there is a short-range attractive
interaction between the particles when the distance is within some
threshold, $d_{ij}<r_{ij}<d_{ij}(1+2u)$.  The parameter $u$ is
introduced to characterize the width ($2u$) and also the strength
($\epsilon u^2$) of the attractive potential. The scale for
  attractive forces is then $\epsilon u/d$. Thus, attractive forces
are characterized by just a single parameter. This greatly reduces the
computational complexity and at the same time keeps the model as
simple as possible.
The inter-particle potential and the corresponding
force is illustrated in Fig.\ref{fig:epot-f}, for a choice
of the parameters $\epsilon,u$.

By choice, we use a simple interaction model which is suitable as a
case-study for the sole effect of switching on the attractive
interactions between particles. In the first approximation, such a
model would be appropriate for attractive emulsion droplets or sticky
grains. In reality, granular materials have much more complicated
interactions. For example, grains interact via frictional forces
\cite{KadauPT2003, GilabertPRE2007, KhamsehPRE2015}.  These are not
considered in this work. The additional complications arising from
frictional interactions, e.g. hysteresis \cite{GrobPRE2014} or
discontinuous shear thickening \cite{ClausPRE2013}, are therefore
excluded for now. We also note that for emulsions, pastes, colloids
etc., the frictionless athermal attractive model provides a good
description.

In addition to the conservative force, a dissipative force acts
between pairs of particles. This viscous force is proportional to
their relative velocity and acts only when particles overlap, i.e.
$r_{ij} < d_{ij}$,
\begin{equation}\label{eq:fdiss}
	\vec{F}_{\text{diss.}} = -b[(\vec{v}_i - \vec{v}_j).\hat{r}_{ij}]\hat{r}_{ij}
\end{equation}
where $b$ is the damping coefficient. In our simulations, $b = 2$
which indicates that our system is overdamped.  We also explore the
rheology for other values of $b$, which we discuss later in the text.

To investigate the rheology of such a system of particles, we
perform molecular dynamics simulations using LAMMPS~\cite{LAMMPS}.

In order to avoid crystallization, we choose a 50:50 binary mixture of
particles having two different sizes, with a relative radii of 1.4. The
system is sheared in $\hat{x}$ direction with a strain rate
$\dot{\gamma}$ using Lees-Edwards boundary conditions. The volume
fraction is $\phi = \sum_{i=1}^{N} \pi R_i^2/L^2$, where $R_i$ is the
radius of the $i$-th particle and $L$ is the length of the simulation
box.  A wide range of volume fractions, from $\phi < 0.50$ to
$\phi=1.0$ has been investigated, and the (repulsive) jamming
transition for this system occurs at $\phi_J \sim 0.8430$. Different
systems sizes have also been studied, viz. $N = 1000$ and $20000$.
Most of the results are reported for $N=1000$.
For the case of analysing the formation of shear-bands, we need
to consider a larger system, for which we use $N=20000$.

The unit of energy is $\epsilon$ and the unit of length is the
diameter of the smaller particle type, $d = 1.0$. The unit of time is
hence $d/\sqrt{\epsilon/m}$, where $m=1.0$ is the mass of the
particles.  The velocity-Verlet algorithm is used to integrate the
particles' equations of motion. All measurements are done, after
steady flow has been reached. Typically, for large $\phi$,
measurements are done over strain intervals of 6, after
a transient initial strain of 2. In the case
of smaller $\phi$, measurements are done over a strain window of $10-15$,
after an initial transient strain of $5-10$. We  ensure steady state conditions, wherein
the observables typically fluctuate  around a constant mean value.

\section{Results}\label{sec:results}

To analyse the effect of weak attractive interactions, we study the
macroscopic physical properties as well as structure and dynamics of
the system at the microscopic level, scanning across a wide range of
densities and attraction strengths.

\subsection{Macroscopic Rheology}\label{sec:macroscopic-rheology}

\subsubsection{Flow Curves}\label{sec:flow-curves}

\begin{figure}[hbt]
	\centering
	\includegraphics[width=0.5\textwidth]{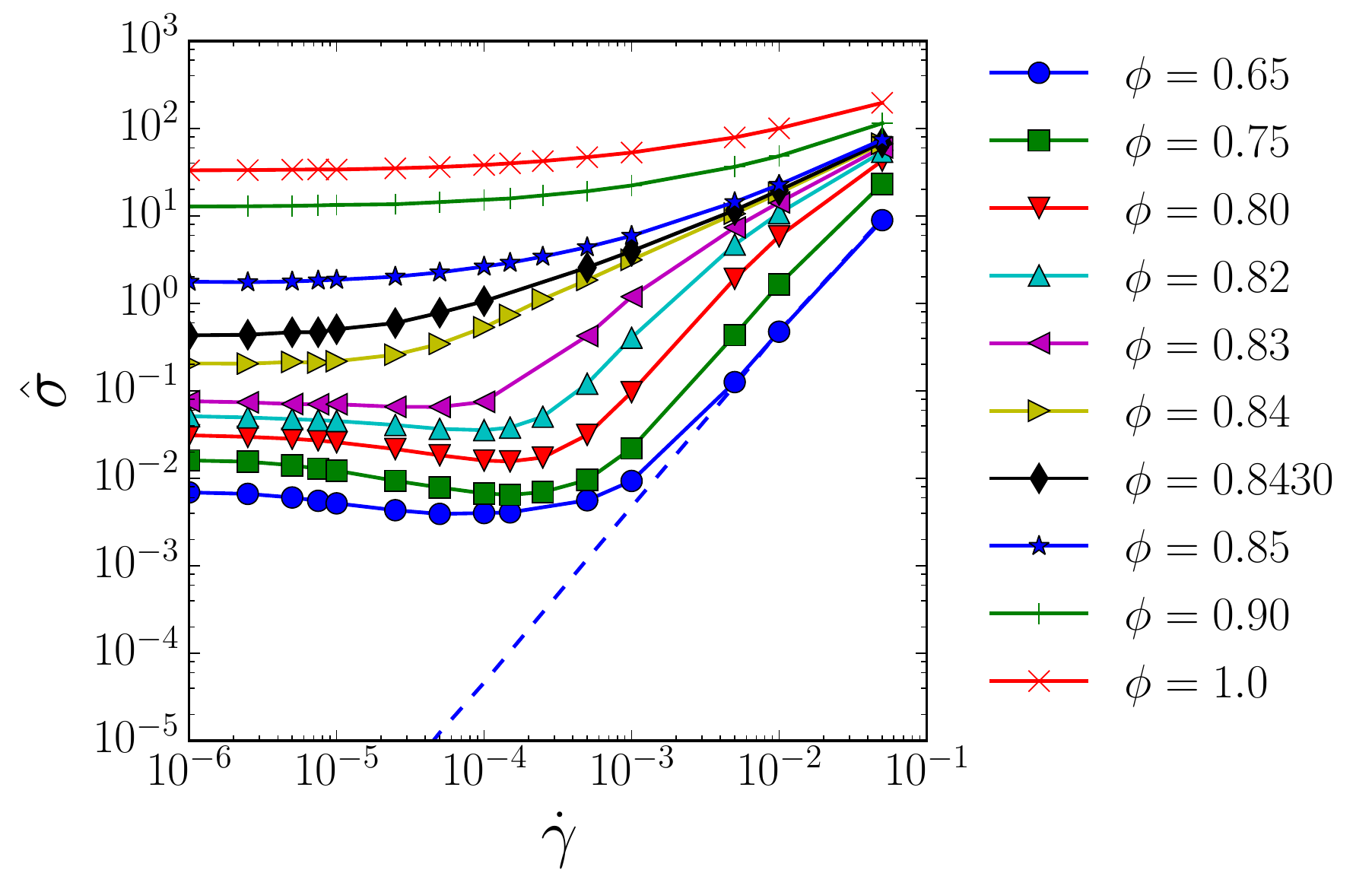}
	\caption{Flow Curves: normalized shear stress
          $\hat\sigma=\sigma d/\epsilon u$ as a function of strain
          rate $\dot{\gamma}$ for different volume fractions. The
          dashed line corresponds to the repulsive system,
          $u=0.0$. Solid lines represent attractive systems with the
          attraction range $u=2\times10^{-4}$. For such systems, the
          jamming transition occurs around $\phi_J \approx
          0.8430$} \label{fig:fc-diffphi}
\end{figure}

The macroscopic rheological response of the system of particles is
characterised by measuring the stress ($\sigma$) that is generated
under the application of external shear rate ($\dot{\gamma}$). When
a system of repulsive particles is sheared at $\phi < \phi_J$,
having the dynamics described, the rheological response shows Bagnold
scaling, i.e. $\sigma \sim \dot{\gamma}^2$.  In Figure
\ref{fig:fc-diffphi}, the dashed line shows such a flow curve for
the repulsive system (with parameters $u=0.0$, $\phi=0.65$ and
$N=1000$).  The question that we address is how such a flow curve
is affected by introducing attractive interactions between the
particles. As shown in our previous work, the material becomes rigid
with the appearance of a yield stress, $\sigma_{\text{y}} =
\sigma(\dot\gamma \rightarrow 0)$, when attraction is added
~\cite{IraniPRL2014}. This can be seen in Figure \ref{fig:fc-diffphi}.
When a weak attraction ($u=2\times10^{-4}$) is switched on, a finite
$\sigma_{\text{y}}$ emerges at $\phi=0.65$.  We also observed that
with increasing $\phi$ at this attraction strength, $\sigma_y$
increases. Thus, for such weakly attractive systems, rigidity sets
in at volume fractions much below the repulsive jamming point
$\phi_J=0.8430$ \cite{IraniPRL2014, SundaresanPRE2014}. Above the
jamming point, flow curves follow a standard Herschel-Bulkley form
which is consistent with earlier results~\cite{PinakiPRE2012}.

The influence of the attractive interactions is observed only in the
regime of small shear-rates. At large shear rates, the dependence of
$\sigma$ on $\dot{\gamma}$ is identical, for the attractive and for
the repulsive systems. One can see this in Figure
\ref{fig:fc-diffphi}, for $\phi=0.65$. Thus, for attractive systems,
we refer to the high shear rate regime as ``repulsion-dominated'' and
the low shear rate regime as ``attraction-dominated''.  In our earlier
work, we have also noted that the range of shear-rates over which the
``attraction-dominated'' regime is observed broadens with increasing
the attraction range and decreasing the volume fraction.

A key feature of the flow curves, in the regime of weak attraction, is
the existence of a non-monotonic shape. We have shown earlier and
rationalised how such a behaviour occurs for $\phi < \phi_J$.  The
non-monotonic dependence in $\sigma$ vs $\dot{\gamma}$ leads to a
mechanical instability, which shows up in the form of localised shear
bands \cite{DhontPRE99, PicardPRE2002}. We will discuss this in more
details in later sections.

\begin{figure}
	\includegraphics[width=0.47\textwidth]{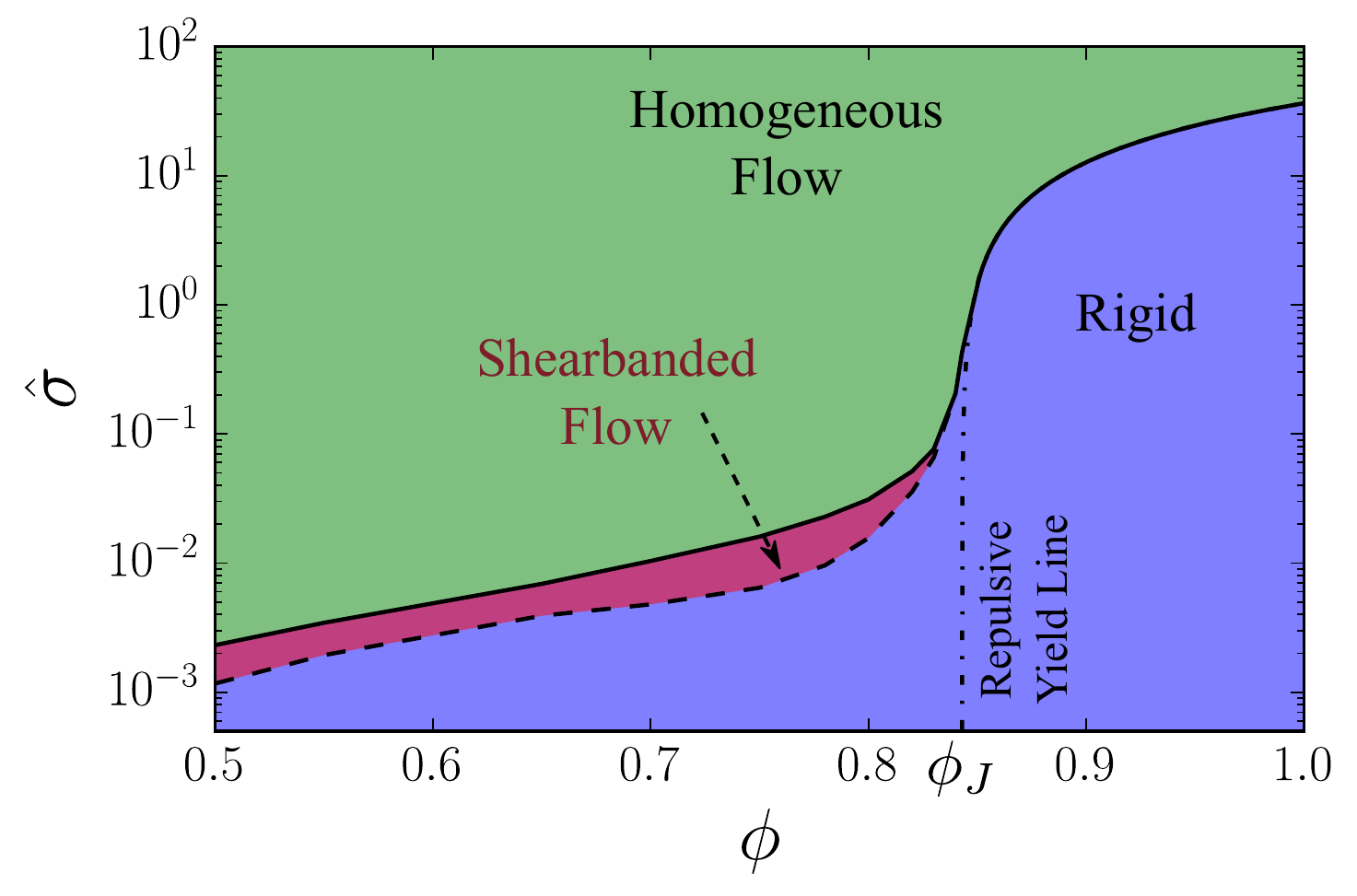}\\
	\includegraphics[width=0.45\textwidth]{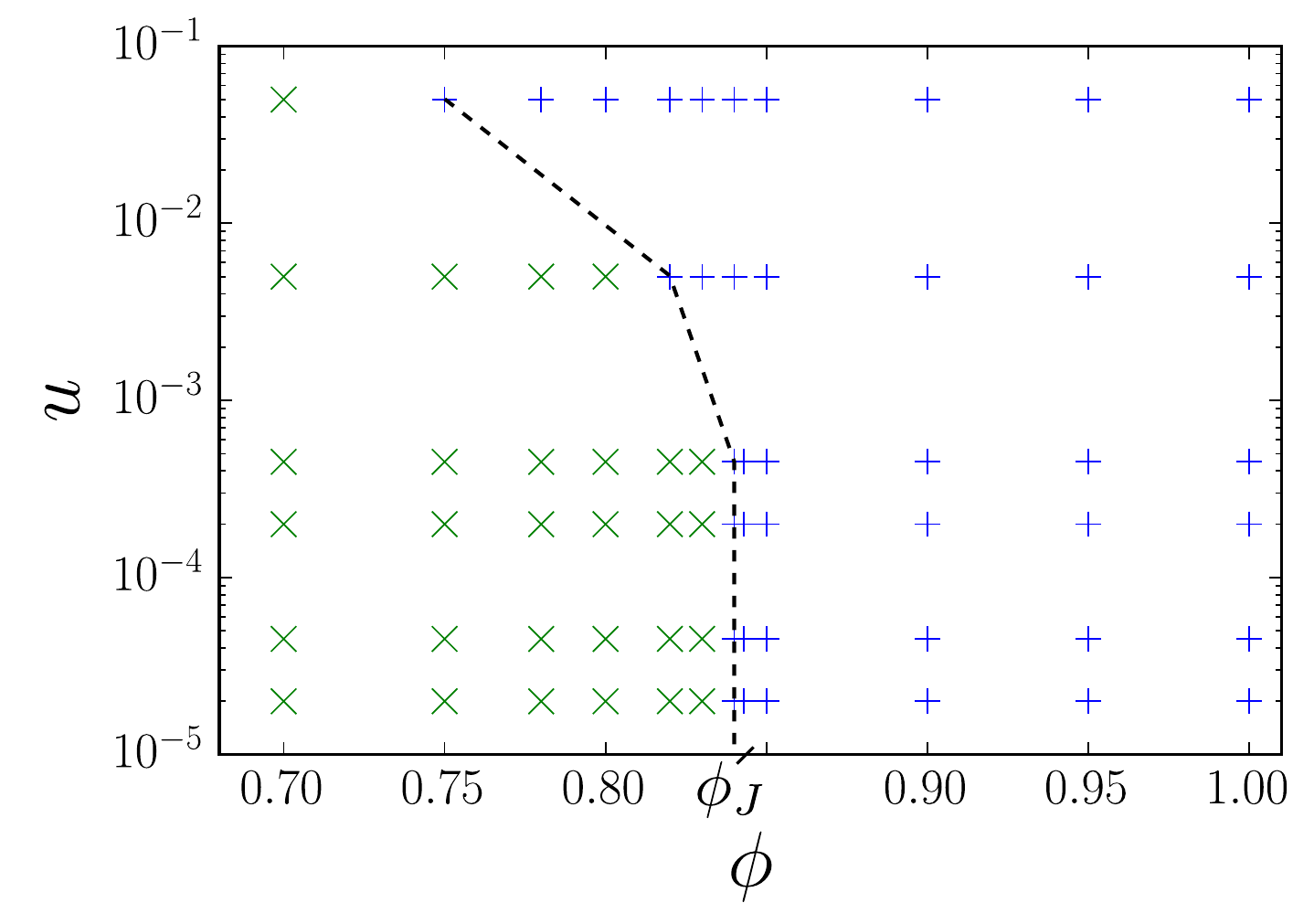}
        \caption{(Color online, top), The modified jamming phase diagram for
          attractive systems, choosing $u=2\times10^{-4}$ : Solid line
          marks the yield stress $\sigma_y$ as a function of
          $\phi$. Above $\phi_J (=0.8430)$, yield stress behaves as
          $\sigma_y \propto \delta{\phi}^{\alpha}$ where
          $\alpha = 1.04$. We also demarcate in the phase diagram, the
          region (shown in purple), where shear-banded flow is
          expected. (Bottom), The state space of $u$ and
          $\phi$, where points to observe shearbands (cross symbols)
          or homogeneous (plus symbols) flow are marked. The critical
          volume fraction of the transition between shearbanded and
          homogeneous flow decreases with increasing the attraction
          strength.}
	\label{fig:phase-diagram}
\end{figure}

For attractive particles, since a finite yield stress is exhibited
even below $\phi_J$, the jamming phase diagram needs to be redrawn.
In the top panel of Figure \ref{fig:phase-diagram}, we show the
necessary modification in the $\sigma-\phi$ plane, for an attraction
strength of $u=2\times10^{-4}$. The solid black line marks the yielding
threshold for this attractive strength, with varying $\phi$. It can
be seen that the system has a finite yield stress even at the
smallest $\phi$ that we have explored, viz. $\phi=0.50$ (see Figure
\ref{fig:phase-diagram}). This is in contrast to the repulsive
system where shear rigidity sets in at $\phi_J$. As indicated in
the figure, for $\phi > \phi_J$, $\sigma_y$ vs $\phi$ is identical
for (weakly) attractive and repulsive systems. We also mark in the
same figure, using dashed lines, the regime in which non-monotonic
flow curves are observed for this attraction strength; for each
$\phi$, this line marks the location of the minimum in the non-momotonic
flow curve. Thus, in this regime, shear-banded steady-state flows
are expected.

The range of attraction over which flow instabilities are observed,
for varying attraction, is indicated in the bottom panel of Figure
\ref{fig:phase-diagram}. We note that with decreasing $\phi$, the
range of attraction strengths over which non-monotonic behaviour
is observed starts to increase and for $\phi=0.7$, such a rheological
response occurs for all attraction strengths that we have explored.

\begin{figure}[htb]
 \centering
 \includegraphics[width=0.45\textwidth]{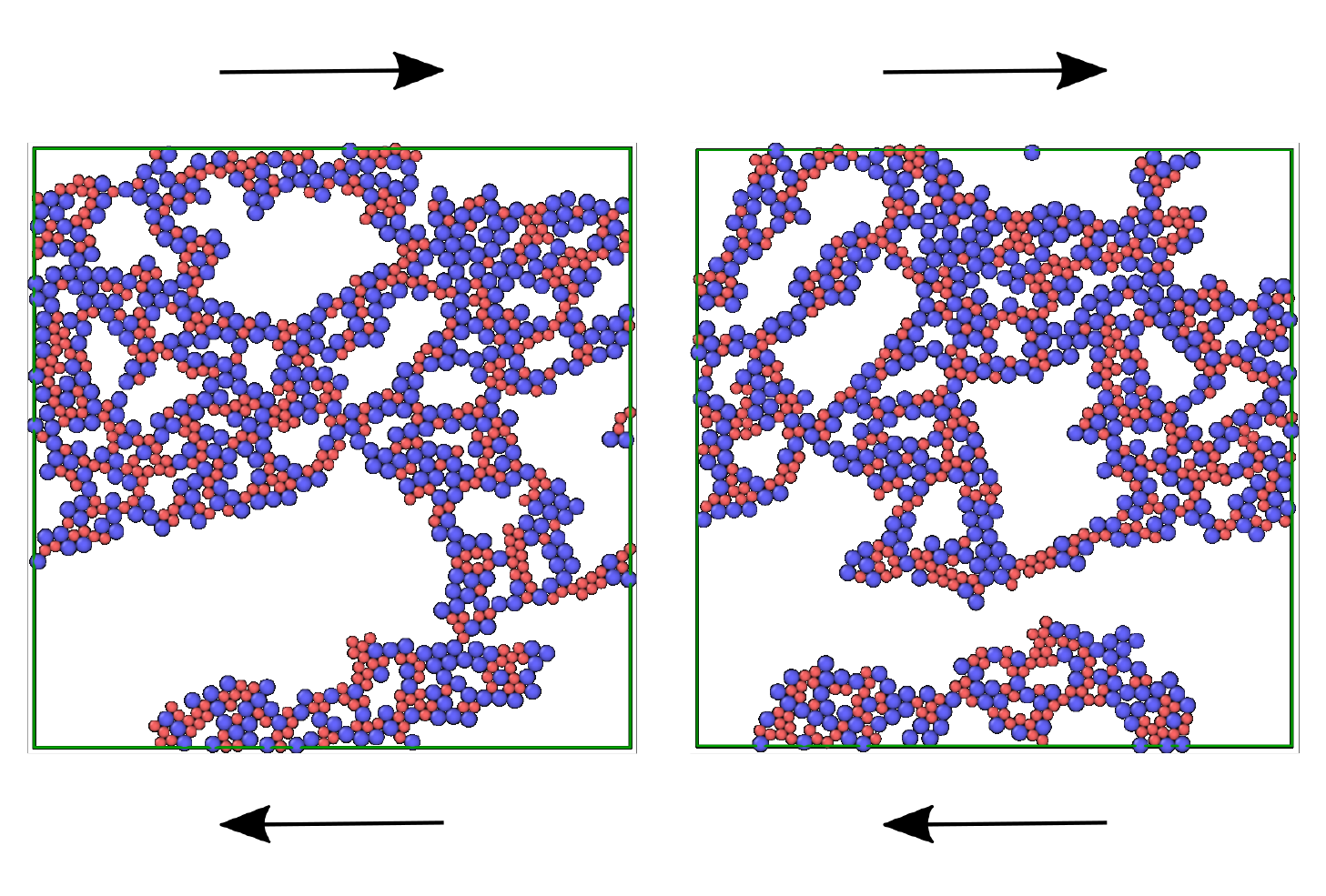}
 \caption{Two snapshots of a dilute system with $N=1000$, $u=2\times10^{-4}$ and
 $\phi=0.40$. (Left), All particles form a large cluster and flow
 due to shearing. (Right), Later in the same system, particles
 form a non-percolating cluster. Arrows
 indicate the shearing direction.} \label{fig:snapshot-low-phi}
\end{figure}

As  is depicted in Figure \ref{fig:phase-diagram} (top), for
attractive particles, a finite yield stress exists even far below
the jamming point where the yield stress vanishes for repulsive
particles. Thus, a valid question is whether there is a threshold
in volume fraction above which a rigidity transition, with a finite
yield stress, occurs for weakly attractive systems.  To answer this
question, one needs to go to smaller $\phi$ than the range depicted
in the phase diagram. However, for $\phi<0.50$, we observe that the
system shows strong sensitivity to the initial configuration and
defining a steady state to measure the yield stress becomes
increasingly difficult. In such systems, particles frequently form
a large cluster and flow for a while, or form two or more smaller
clusters which do not interact most of the time, due to the large
void spaces between them; see Fig. \ref{fig:snapshot-low-phi}.
Therefore the stress measurements become very sensitive to the
strain window over which the measurement is performed. Therefore,
in this work, we focus only on relatively denser systems ($\phi\ge0.50$)
where situations as depicted in  Fig.~\ref{fig:snapshot-low-phi}
do not occur and a steady state is  absolutely well defined.

\subsubsection{Coordination Number}\label{sec:coordination-number}

The flow response is related to the underlying contact network formed
by the particles. The structure of the network is characterised by the
connectivity, $z$, which is the average number of contacts per
particle.
To measure $z$, we count all neighbors in the range of the interaction
potential.  Thereafter, we divide these contacts into two different
types. When $r_{ij}{\leq}d_{ij}$, i.e. the particles repel each other,
we speak of ``repulsive contacts''. With the inclusion of attractive
forces, we also define attractive contacts ($z_{\text{att}}$) when
$1<r_{ij}/d_{ij}<1+2u$. The total coordination number is defined as
$z=z_{\text{rep}}+z_{\text{att}}$.

\begin{figure}[htb]
	\centering
	\includegraphics[width=0.45\textwidth]{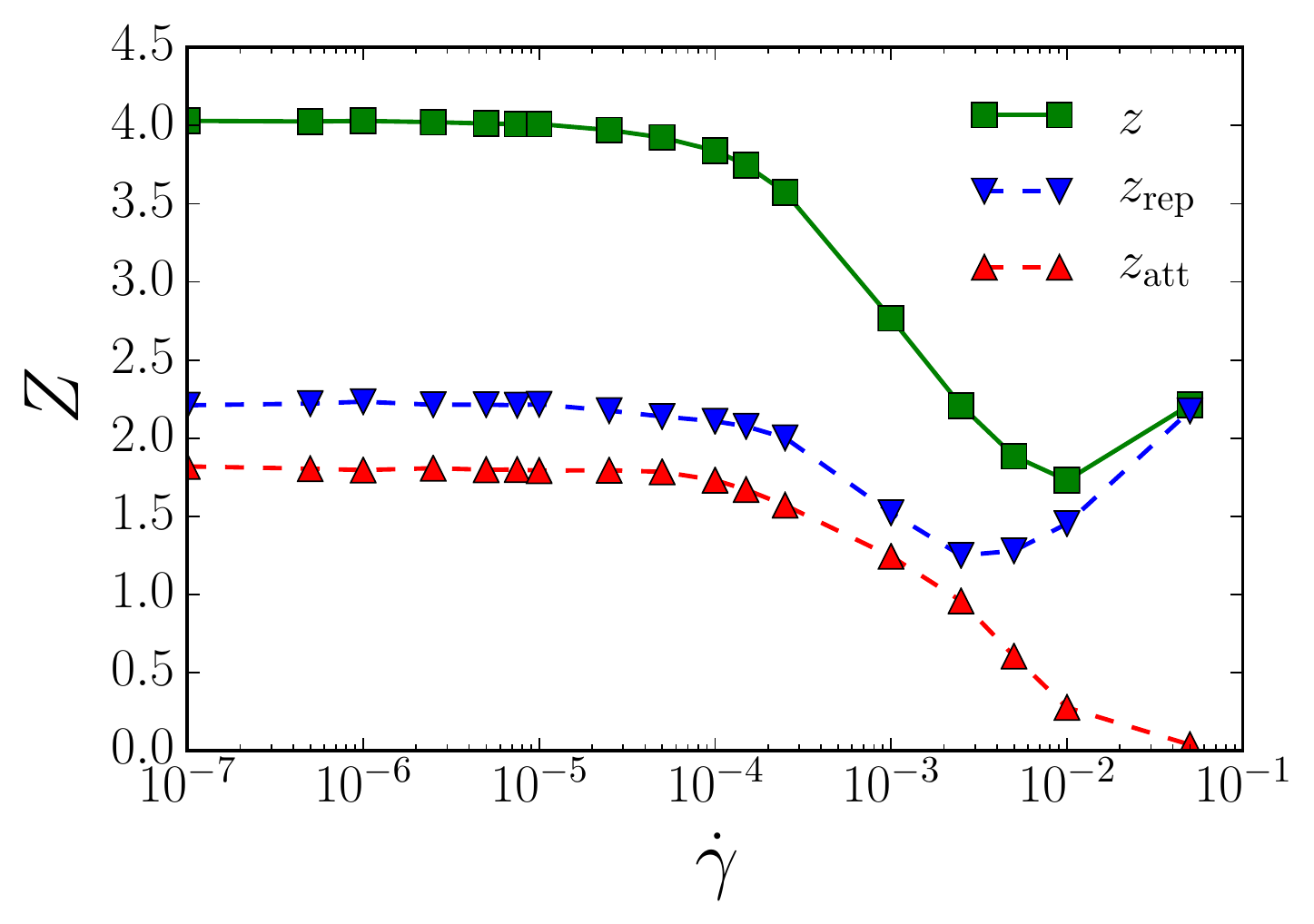}
	\caption{Variation of different kinds of connectivity, $z,  z_{\text{rep}}, z_{\text{att}}$,
	as defined in the text, with strain
	rate for a system at $\phi=0.75$ and $u=2\times10^{-4}$. At small strain
	rates, attractive contacts play important role and keep the
	total connectivity larger than the isostactic value. at large
	strain rates, repulsive contacts are dominant, determining
	the total connectivity behaviour. The drop in $z(\dot\gamma)$
	from isostactic point to the value much smaller, indicates the
	attractive timescale $\tau_a$.} \label{fig:z_fc}
\end{figure}

In Figure \ref{fig:z_fc}, for a fixed value of $u$, we show how
$z_{\text{rep}}$, $z_{\text{att}}$ and $z$ vary with shear-rate, for
an attraction strength where the flow curve is non-monotonic. The data
is taken at $\phi=0.75$, far below $\phi_J$. At small strain rates,
both $z_{\text{rep}}$ and $z_{\text{att}}$ are constant, with slightly
more repulsive contacts than attractive ones. With increasing strain
rate, the attractive contacts rapidly decay and become negligible at
large shear-rates. On the other hand, repulsive contacts exhibit a
non-monotonic behaviour. While contacts initially get disrupted and
thereby decrease with increasing shear-rate, the particles once again
get pushed together when the repulsive regime kicks in.  Thus, at
large shear-rates, $z$ is entirely dominated by the repulsive
contacts.

\begin{figure}[htb]
	\centering
	\includegraphics[width=0.45\textwidth]{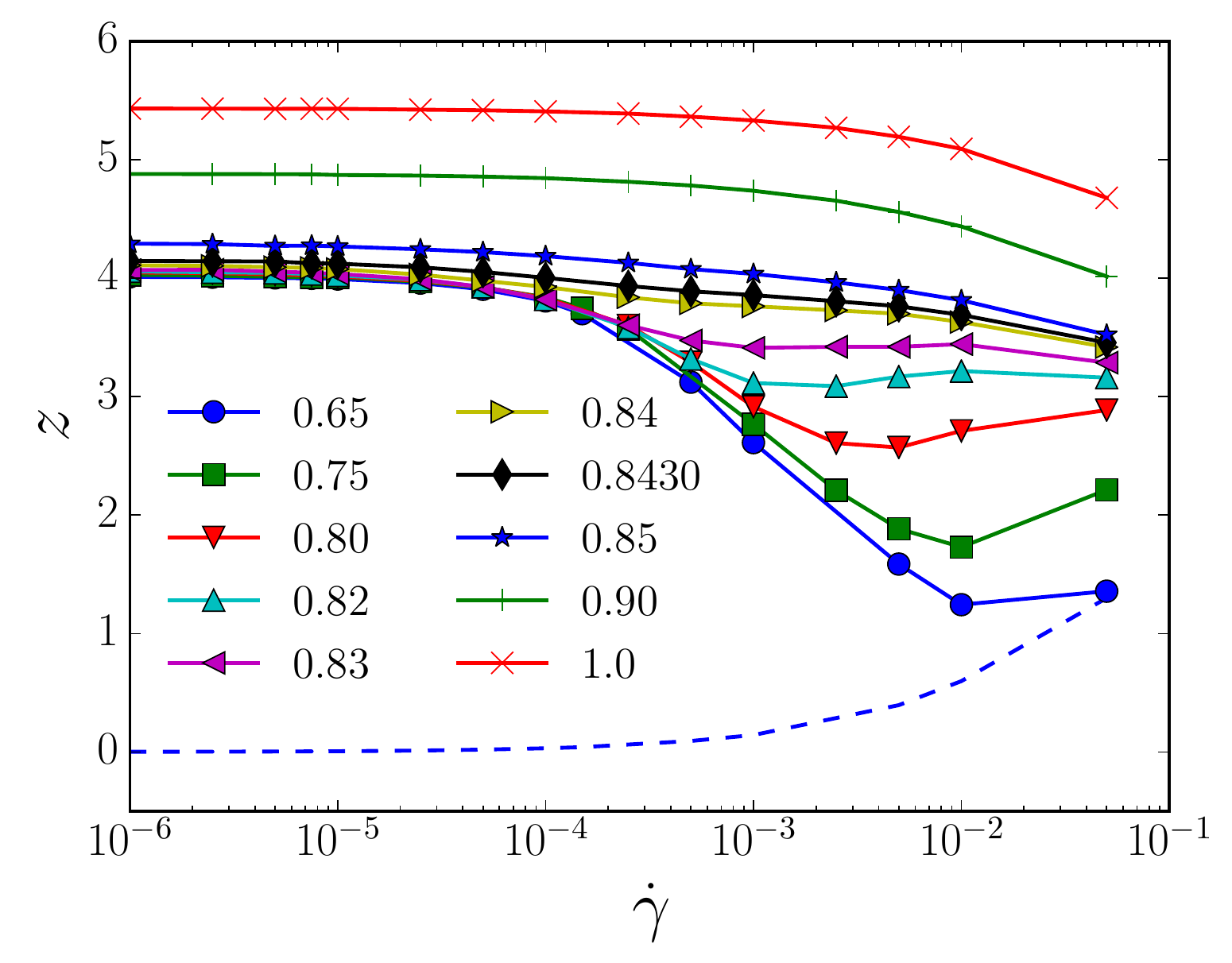}
	\caption{$z(\dot\gamma)$ for different volume fractions.
	Similar to Fig. \ref{fig:fc-diffphi}, the dashed line
	corresponds to the repulsive system $u=0.0$, solid lines
	represent attractive system with $u=2\times10^{-4}$.}
	\label{fig:z_fc_diff_phi}
\end{figure}

In Figure \ref{fig:z_fc_diff_phi}, we plot $z(\dot\gamma)$ for
different $\phi$. The dashed line represents the connectivity in a
repulsive system at $\phi=0.65$ and the solid lines corresponds to the
attractive systems with $u=2\times10^{-4}$. It is observed that
$z(\dot\gamma) \rightarrow 0$ at $\dot\gamma \rightarrow 0$ for
repulsive particles below jamming, which is expected for our model of
particles dynamics \cite{TeitelPRL2014}. On the other hand, for
attractive systems and below jamming, as soon as the attraction is
introduced, $z_y \equiv z(\dot\gamma \rightarrow 0)$ jumps to a value
slightly larger than the isostatic connectivity
$z_{\mathrm{iso}} = 4$.  Consistent with the observation of Khamesh et
al \cite{KhamsehPRE2015}, no rattlers occur.  Figure
\ref{fig:z_fc_diff_phi} reveals that such a behavior holds for $\phi$
even far below $\phi_J$. Therefore, finite but small attraction for
$\phi < \phi_J$ results in similar isostactic structures as the
structure for $\phi=\phi_J$ in repulsive systems.  The threshold value
$z=4$ is to be understood as mean-field result and neglects the
possibility of mechanisms and states of self stress. There is no
reason to believe that these are absent in our networks. However, in
line with other studies in the field these effects do seem to play
only a minor role~\cite{vanHeckeJPCM2010}.

For $\phi>\phi_J$ we find $\delta z \equiv z - z_{\mathrm{iso}}=\zeta_0
(\phi-\phi_J)^{1/2}$ with $\zeta_0\approx3.78$,  consistent
with~\cite{OHernPRE2003}. In the limit of large strain rates,
$z(\dot\gamma)$ reproduces the repulsive results as it is shown for
the system at $\phi=0.65$.

As one can see in Figures \ref{fig:z_fc_diff_phi} and \ref{fig:z_fc},
in the limit of small strain rates, $z(\dot\gamma)>z_{\text{iso}}$.
It remains almost constant with $\dot\gamma$ until at a special strain
rate, it drops to values far below $z_{\text{iso}}$. This behaviour of
connectivity indicates that at small strain rates, attractive forces
keep the nearly-isostactic structure of the system stable until the
point where the rate of deformation is fast enough to destroy the
structure. This rate gives us an attractive time-scale $\tau_a$ which
is found to scale with the attraction range as $\tau_a \sim 1/u$
independent of $\phi$~\cite{IraniPRL2014}.

Recognition of such a timescale also helps in rationalising the
variation of stress with shear-rate, as shown in
Figure~\ref{fig:fc-diffphi}.  At small $\dot\gamma$, the relaxation
time is much smaller than the shearing time scale which leads to a
continuous reconstruction of the network structure. On the other hand
at large $\dot\gamma$ this local structure breaks down due to fast
shearing and a relatively large relaxation time. The competition
between these two mechanisms, attraction induced aggregation and shear
induced rupture of local structure, explains the decrease of stress in
the intermediate regime.

\begin{figure}[hbt]
	\centering
	\includegraphics[width=0.40\textwidth]{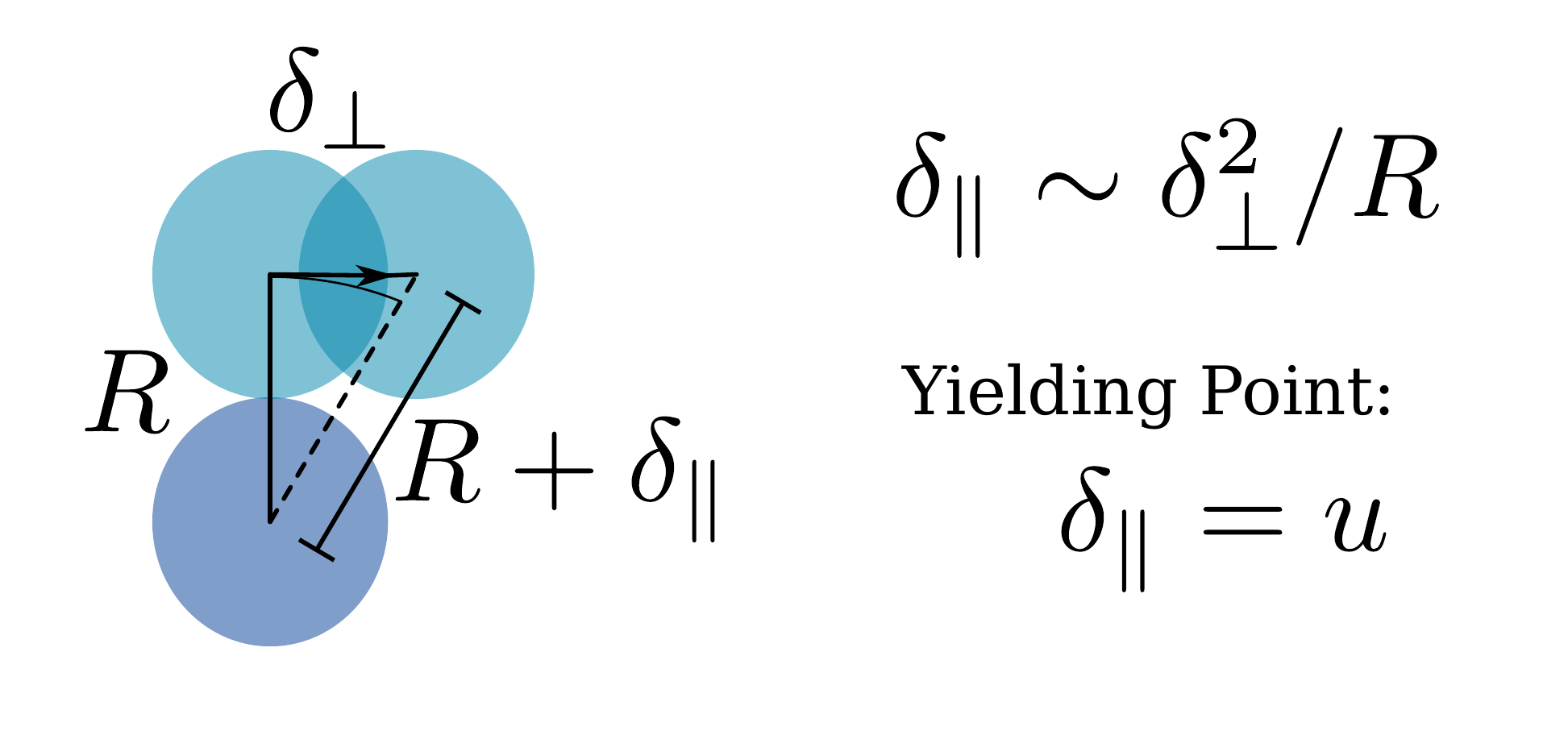}
	\caption{Sketch of a pair of particles at the yielding point.}
	\label{fig:particles-disp}
\end{figure}

Now we focus on how the yield stress is related to the connectivity of
the system below the jamming point. In elastic spring networks and
systems of soft repulsive particles it is well
known~\cite{vanHeckeJPCM2010} that the linear response to a
macroscopic shear strain $\gamma$ close to the isostatic point is
characterized by strong non-affine motion, expressed by relative
particle displacements $\delta_\bot\sim \gamma/\delta z^{1/2}$. Those
displacements are directed tangentially to the particle contact (see
Fig.~\ref{fig:particles-disp}). The corresponding shear modulus of
such a linear-elastic response is $g_{\text{lin}} \sim \delta z$. In
our system when the contact is not broken, particles see each other
through the harmonic force with a range determined by $u$. Thus for
motion amplitudes smaller than this range, our system can be
considered as a network of elastic springs and yielding can be defined
as the point where the motion amplitude is larger than the range of
attraction and breaks the contact.

To determine the yielding point, we need non-linear loading conditions
where the particle's tangential motion is associated with higher-order
longitudinal contributions, $\delta_\parallel \sim \delta_\bot^2/R$
(Pythagoras) (similar reasoning has been applied in different
contexts, see \cite{ClausPRL2007,WyartPRL2008}). In our attractive
system, the maximum dilational strain should be smaller than the
attraction range, $\delta_\parallel < u$ (Figure
\ref{fig:particles-disp}). The yield strain can then be expressed as
$\gamma_{\text{y}} \sim (u\delta z)^{1/2}$. Since the yield stress is
$\sigma_{\text{y}} \sim g_{\text{lin}} \gamma_{\text{y}}$, one can
write the following scaling relation for the yield stress, attraction
range and the distance to the isostatic point:

\begin{equation}
\label{eq:att-yield-stress}
 \sigma_{\text{y}} \sim u^{1/2}\delta z^{3/2}
\end{equation}

\begin{figure}[hbt]
	\centering
	\includegraphics[width=0.50\textwidth]{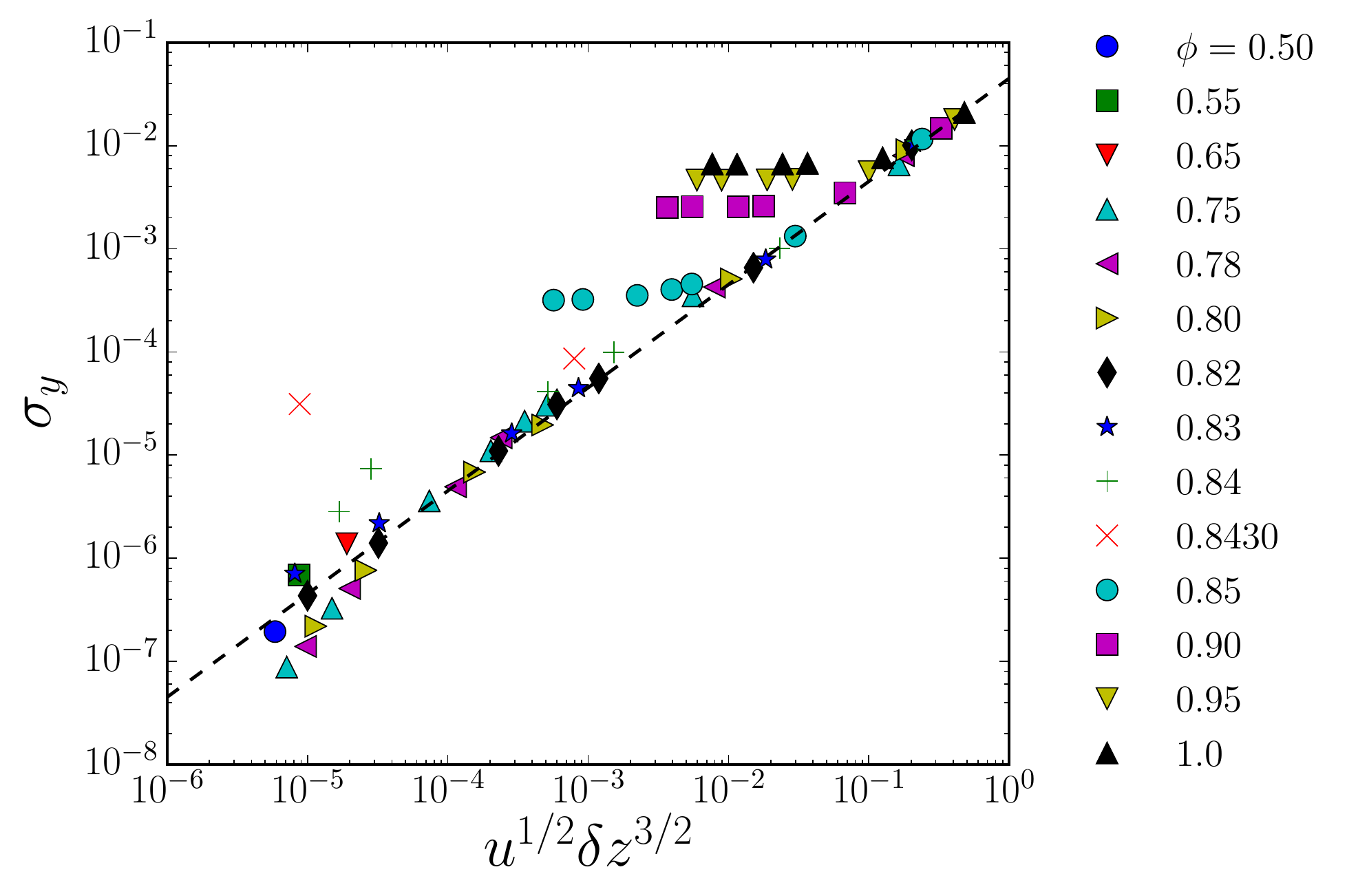}
	\caption{Yield stress $\sigma_y$ as a function of the
          combination $u^{1/2}\delta z^{3/2}$. For small $\phi$ or
          $u$, Eq. \ref{eq:att-yield-stress} holds (dashed line). For
          large $\phi$ or $u$, the yield stress shows the expected
          repulsive behaviour $\sigma_y \sim |\delta \phi|^\alpha$.}
	\label{fig:yield-stress-scaling}
\end{figure}

The weak attractive forces we use in our model result in the formation
of a fragile solid (nearly-isostactic network of particles). The
mechanical response of this fragile solid is explained by Equation
\ref{eq:att-yield-stress}. Figure \ref{fig:yield-stress-scaling} shows
that Equation \ref{eq:att-yield-stress} holds nicely for systems with
a wide range of volume fractions below and above the jamming point and
different attraction ranges (Dashed line). Above the jamming point and
at high enough $u$, the well-known repulsive behaviour $\sigma_y \sim
|\delta \phi|^\alpha$ is observed. It can be
understood by noting the fact that in highly dense systems, the
repulsive term of Equation \ref{eq:pot} is dominant.

\subsubsection{Potential Energy}\label{sec:potential-energy}

\begin{figure}[h]
	\centering
	\includegraphics[width=0.5\textwidth]{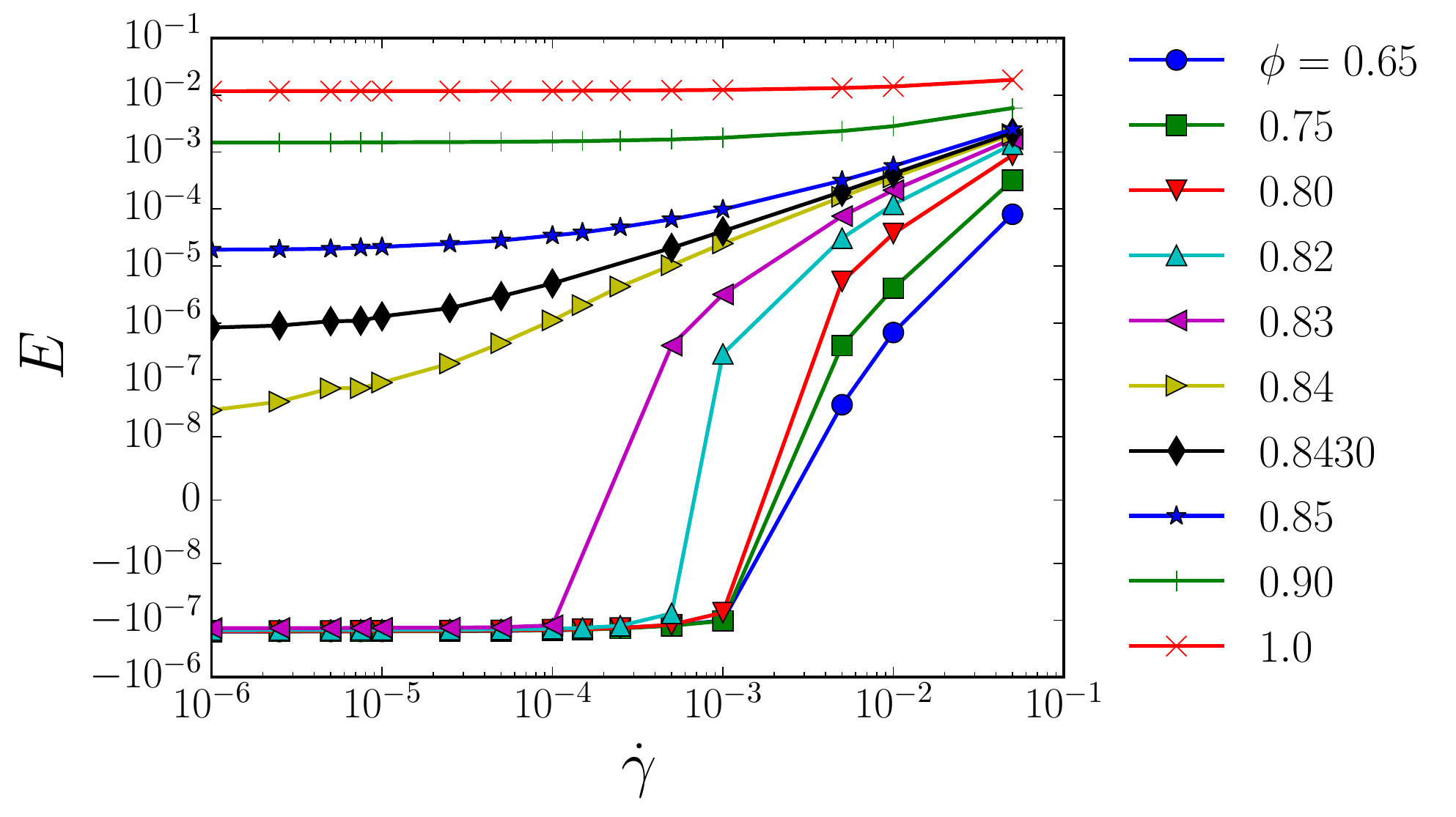}
	\caption{Potential energy per particle $E$ as function of
          strain rate $\dot{\gamma}$ for different volume
          fractions. The dashed line represents the repulsive system
          with vanishing energy in the limit of zero strain
          rate. Solid lines are associated with attractive systems
          with the attraction range
          $u=2\times10^{-4}$}.\label{fig:fc-epot}
\end{figure}

The attractive interactions introduce a new energy scale,
$\epsilon u^2$.  We now investigate how this shows up in the
rheology. In the top panel of Figure \ref{fig:fc-epot}, we plot how
the potential energy per particle, $E$, varies with imposed
shear-rate, for various $\phi$, using $u=2\times10^{-4}$. For
$\phi > \phi_J$, the potential energy per particle is dominated by
repulsive contributions and is thus positive.  For $\phi < \phi_J$, at
small $\dot{\gamma}$, $E$ becomes negative as attractive forces become
responsible for stabilising the solid. However, with increasing
$\dot{\gamma}$, again the particles come into physical contact and the
energy becomes positive. In fact, one can see that at $\phi=0.65$, for
large $\dot{\gamma}$, the potential energy per particle for these
attractive particles is the same as for the repulsive particles (shown
with dashed lines). Since the average number of contacts is $z$, the
variation of the potential energy per particle due to the attractive
bonds with changing shear-rate can be written as

\begin{equation}
E=-\epsilon u^2z(\dot\gamma).
\label{eq:attractive-epot}
\end{equation}

Earlier, in Figure \ref{fig:z_fc}, we had seen that $z(\dot\gamma)$ is
almost constant in the small $\dot{\gamma}$ regime, below
$\phi < \phi_J$. Furthermore, it also does not vary much with $\phi$
(see Figure \ref{fig:z_fc_diff_phi}). Thus, in this regime, the
potential energy per particle is also expected to be constant, which
is illustrated in the bottom panel of Figure \ref{fig:fc-epot} for
different $\phi$.

\begin{figure}[htb]
	\centering
	\includegraphics[width=0.5\textwidth]{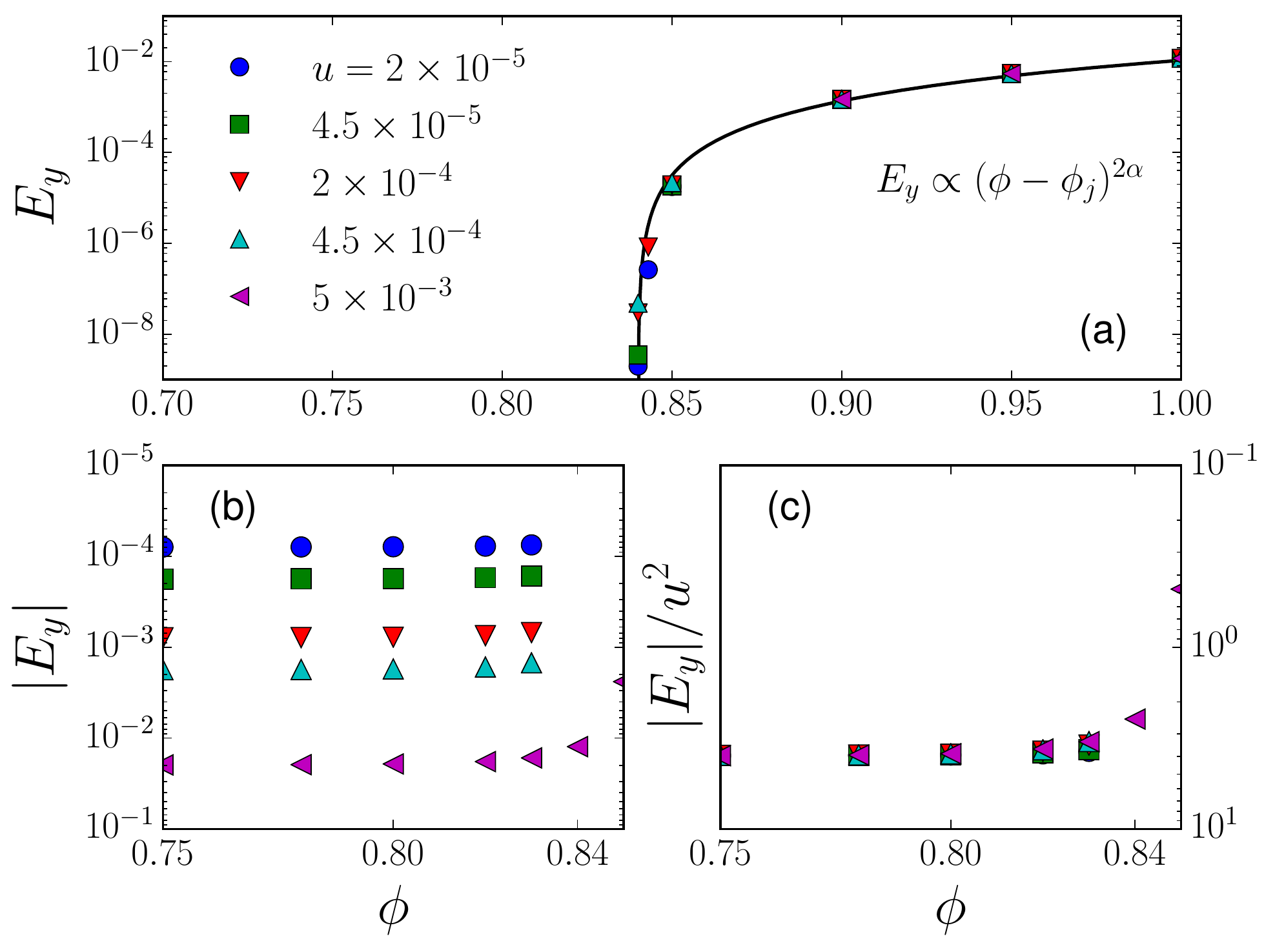}
	\caption{Yield potential energy $E_y$ as a function of $\phi$
	for different attractions $u$ in (a) repulsion-dominated
	regime, $\phi>\phi_J$ and (b) attraction-dominated regime,
	$\phi<\phi_J$. (c) Scaled $E_y(\phi)$ by $u^2$ in the
	attraction-dominated regime.} \label{fig:epot-y-phi-scaled}
\end{figure}

Similar to defining a stress threshold for yielding to occur, one can
define a yield potential energy as $ E_y = E(\dot{\gamma} \rightarrow
0)$. The variation of the estimated $E_y$ with $\phi$ for different
attraction strengths is shown in Figure
\ref{fig:epot-y-phi-scaled}. For $\phi > \phi_J$, where repulsion
dominates, we see that $E_y \propto (\phi - \phi_J)^\beta$ where we
find $\beta \approx 2.1$. For repulsive particles, $E \propto
\sigma^2$, implying that $\beta = 2\alpha$ which is consistent with
our observations. For $\phi < \phi_J$, from Equation
\ref{eq:attractive-epot}, one expects $|E_y| \sim u^2$, and this is
demonstrated in Figure \ref{fig:epot-y-phi-scaled}(c).

\subsubsection{Shear Stress Ratio}\label{sec:shear-stress-ratio}

The isostatic solid formed due to attractive interactions, at
$\phi<\phi_J$, is resistant to plastic deformations under shear. In
granular mechanics, this macroscopic resistance (or friction) is
quantified via the ratio of shear-stress to pressure:
\begin{equation}
 \mu={\sigma_{\text{xy}} \over P}.
\end{equation}

\begin{figure}[hbt]
	\centering
	\includegraphics[width=0.5\textwidth]{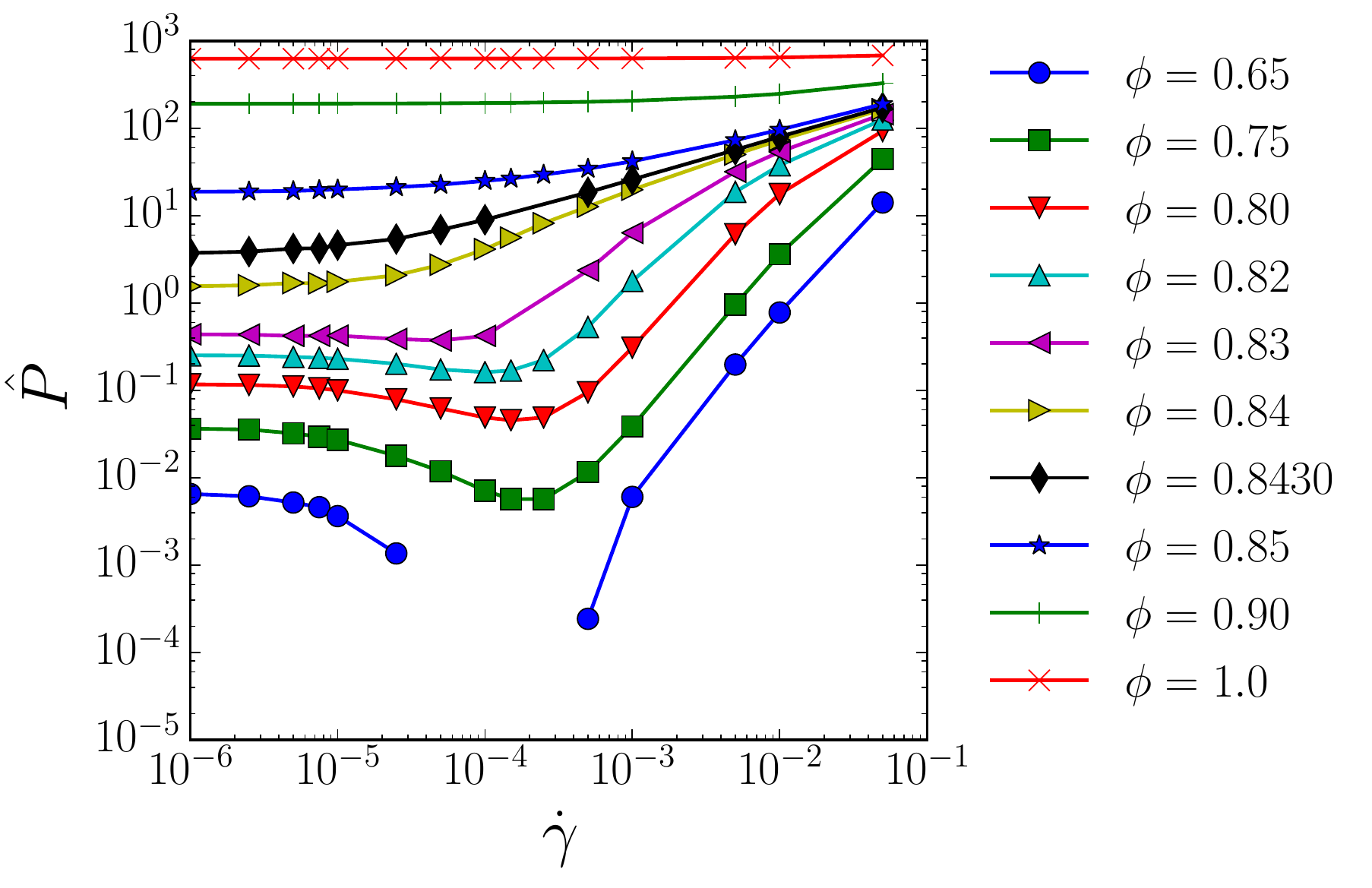}
	\caption{Rescaled pressure $\hat P=P d/\epsilon u$ as a
          function of strain rate $\dot\gamma$ for different volume
          fractions. The attraction range is $u=2\times10^{-4}$. The
          discontinuity in $P(\dot\gamma)$ at $\phi=0.65$ corresponds
          to negative values of $P$ (compressive pressure). }
	\label{fig:fc-press-diff-phi}
\end{figure}

To aid our discussion on the macroscopic resistance, we also show
the data for pressure, using a rescaled form, in Figure
\ref{fig:fc-press-diff-phi}, for the same set of $\phi$ for which
shear-stress is shown in Figure \ref{fig:fc-diffphi}. The non-monotonic
shape of the pressure curves look similar to that of shear stress,
the quantitative comparison of which is captured by $\mu$, discussed
below. The striking observation is that at $\phi=0.65$ the pressure
curve becomes discontinuous on the log-scale. In that window the
pressure is actually negative, implying the dominance of internal
tensile forces.

\begin{figure}[htb]
	\centering
	\includegraphics[width=0.5\textwidth]{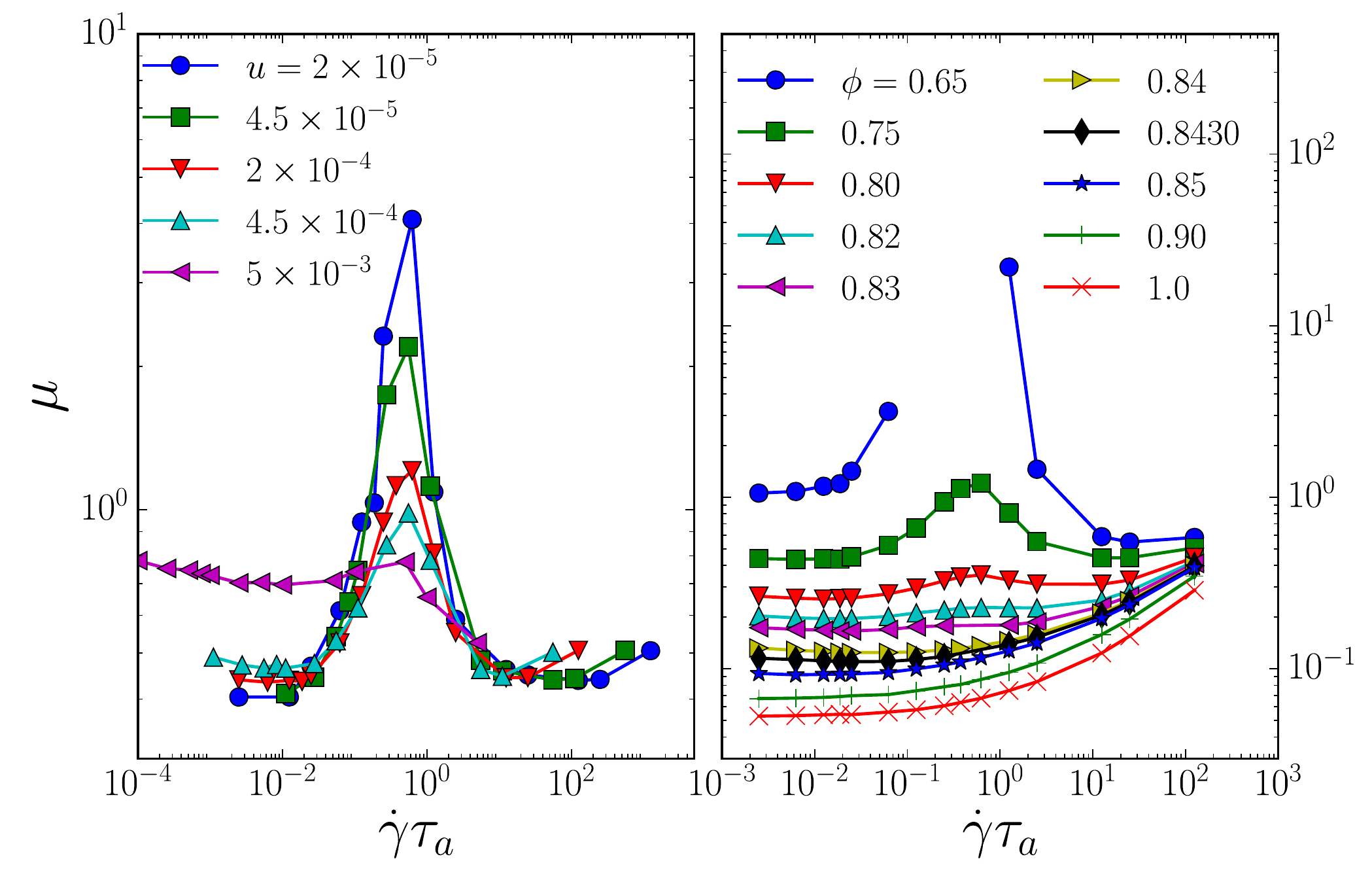}
	\caption{Shear-stress ratio as a function of $\dot\gamma$ for
          different attraction ranges at $\phi=0.75$ (left panel) and
          for different volume fractions at $u=2\times10^{-4}$. Note
          that for $\phi<\phi_J$, the peak appears close to
          $\dot\gamma^*$ where $\sigma(\dot\gamma^*)$ is minimum.}
	\label{fig:stress-ratio}
\end{figure}

Now, we check how the macroscopic friction $\mu$ varies with strain
rate scaled by the attractive time-scale, $\dot\gamma\tau_a$.  In the
left planel of Fig.\ref{fig:stress-ratio} we show the variation of
$\mu$ for a fixed volume fraction ($\phi=0.75$) and different
attraction strengths. The right panel in Figure \ref{fig:stress-ratio}
shows $\mu$ for a fixed attraction ($u=2\times10^{-4}$) but different
volume fractions. Both panels highlight the fact that in the presence
of finite attraction and for $\phi < \phi_J$, $\mu$ develops a peak at
the strain rate close to $\dot\gamma^*$, where the minimum occurs in
flow curves, $\sigma^{\text{min}}=\sigma(\dot\gamma^*)$. In fact, due
to the rescaling of $\dot{\gamma}$ by $\tau_a$ the location of the
peak is at the same $\dot\gamma^*$ for different $u$, with the peak
height decreasing with increasing attraction, which is in agreement
with earlier studies \cite{RognonRouxFLM2008, SundaresanPRE2014}.  In
the right panel, for $\phi < \phi_J$, this non-monotonic behaviour is
observed to disappear for $\phi > \phi_J$. Recalling the discussion
above, the apparent discontinuity of $\mu$ for $\phi=0.65$ is due to
the pressure becoming negative. The observed non-monotonic behaviour
of $\mu$ for $\phi < \phi_J$ happens because, under shear, the
pressure of the particle assembly drops faster than the shear
stress. This also leads to changes in microstructure, which we discuss
in a later section.

Also in order to locate our analysis with other works on cohesive
grains \cite{BergerEPL2015,KhamsehPRE2015,SinghPRE2014}, we have computed the inertial
number $I=\dot\gamma\sqrt{m/P}$, where $m$ is the mass of the
particles and $P$ is the pressure. We find that non-monotonic
flowcurves occur for small values of $I<0.05$, in agreement with
\cite{KhamsehPRE2015}. Also the values of the rescaled pressure
$\hat P = O(1)$, where non-monotonic effects first occur (see
Fig.~\ref{fig:fc-press-diff-phi}), agree with the previous findings
(Ref. \cite{KhamsehPRE2015}, using longer range attractive interactions,
finds shear bands for $\hat P<0.1$).



\subsubsection{Role of damping}\label{sec:role-damping}

\begin{figure}[htb]
\centerline{\includegraphics[width=0.45\textwidth]{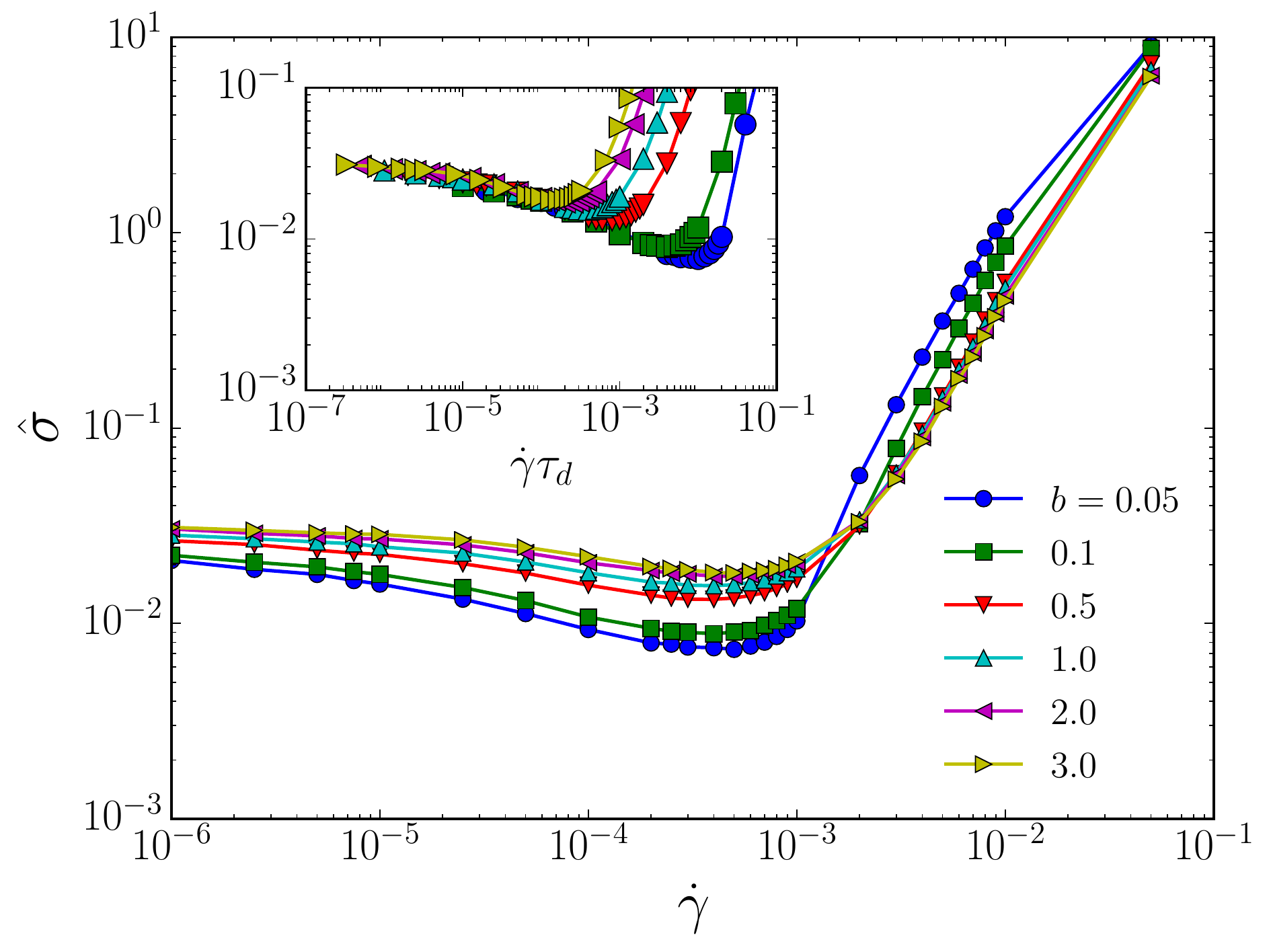}}
\caption{ (Main) Flow curves ($\hat\sigma=\sigma d/\epsilon u$), for
  $\phi=0.75$ and $u=7\times10^{-4}$, with changing damping
  coefficient $b$.  (Inset) The low shear-rate regime of the flow
  curves for different $u$ can be collapsed by using the rescaled
  variable $\dot{\gamma}\tau_d$, where $\tau_d=m/b$.}
\label{fig:damping}
\end{figure}

Our main focus is modelling soft athermal materials with attractive
inter-particle interactions, like in emulsions or gels. In such cases,
the motion of the constituent particles is over damped. Thus, in this
manuscript, we have reported results for such damped systems, with a
relatively large damping parameter $b=2$. Recent studies have explored
how the rheological response changes as one tunes the dynamics from
being over-damped, as is the usual case for suspensions, to being
under-damped \cite{SalernoPRE2013, NicolasPRL2016}. We now explore
this scenario by tuning the damping parameter $b$. The results are
seen in Fig.~\ref{fig:damping}.

\begin{figure}[hbt]
\centering{\includegraphics[width=0.5\textwidth]{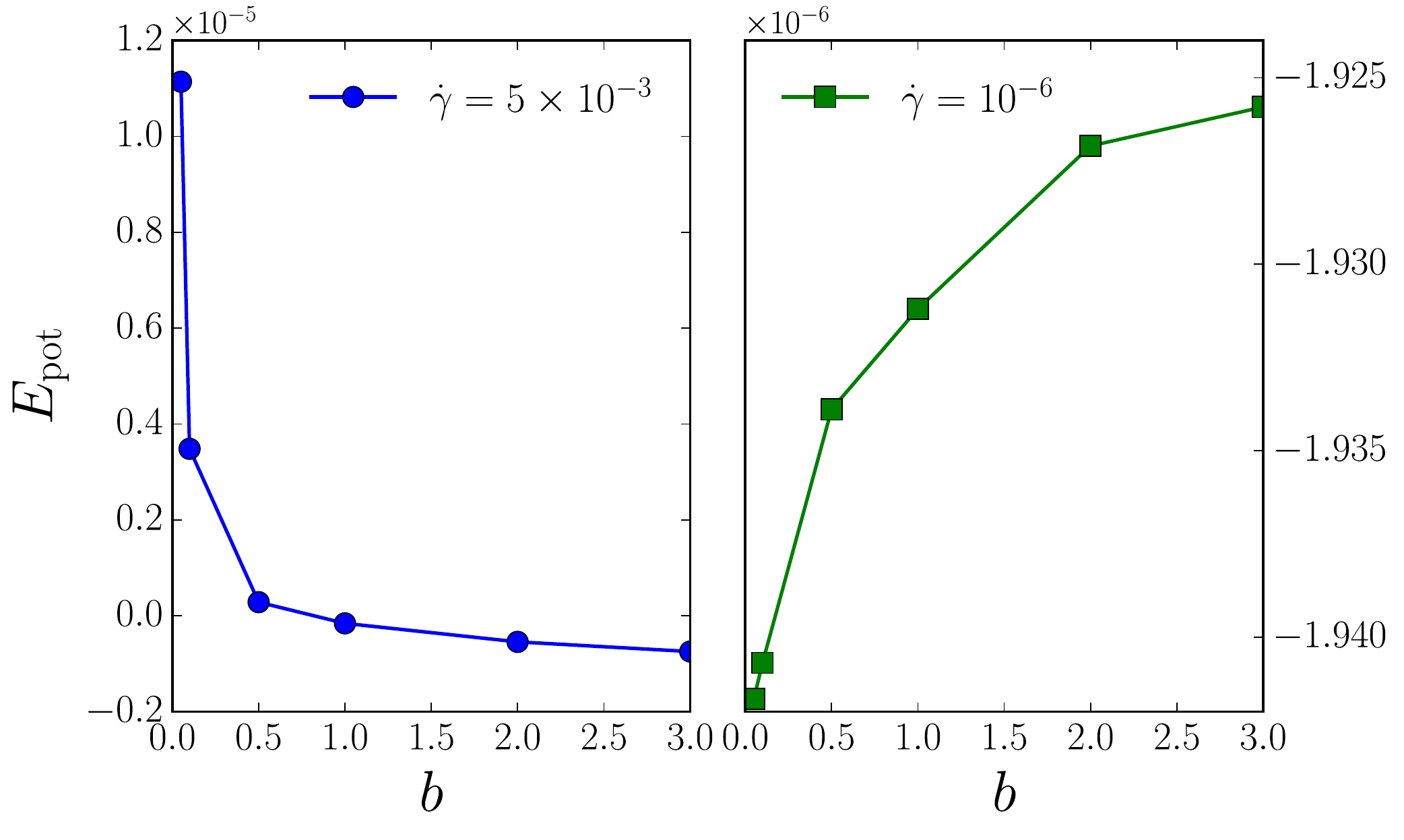}}
\caption{For $\phi=0.75$, $u=7\times10^{-4}$, variation of the
  potential energy per particle $E$ with damping factor $b$,
  (left) in the repulsion-dominated regime with
  $\dot\gamma=5\times10^{-3}$ and (right) in the
  attraction-dominated regime with $\dot\gamma=10^{-6}$.}
\label{fig:epot-b}
\end{figure}

We observe that with decreasing the damping coefficient the
non-monotonicity in the flow curves become more pronounced. Thus,
underdamping enhances the mechanical instability in the system. In
effect, by decreasing the damping coefficient, the timescale for
energy dissipation increases. This leads to the system being in a
fluidized state up to higher strain rates. This can be seen if we
rescale the imposed shear-rate $\dot{\gamma}$ with the dissipation
timescale defined as $\tau_d=m/b$ (inset of
Fig.~\ref{fig:damping}). Thereby, the data for different $b$ can be
collapsed in the low shear-rate regime, with the tuning of $b$ leading
to the exploration of different regimes along this branch. We have
also observed that if we look at the variation of $E$ with changing
$b$, there is a change with decreasing $\dot{\gamma}$.  Typically, at
large $\dot{\gamma}$, the potential energy per particle of the system
decreases as the system gets more damped (i.e. increase of $b$);
quicker dissipation leads to the system not being able to explore all
possible higher energy states.  However, for $\dot{\gamma}$ in the
"attractive regime", we see that $E$ decreases with decreasing $b$;
the underdamping allowing the system to explore lower energy states in
the landscape (Fig.~\ref{fig:epot-b}). Such a scenario has also been
proposed in Ref.\cite{ NicolasPRL2016} to understand how damping
influences steady state rheology of amorphous systems and our
observations are consistent with that.

\subsection{Structure and dynamics}\label{sec:structure-dynamics}

Next, in order to understand the properties of the system in the
different flow regimes, we study the structure and arrangement of
particles as well as their dynamics.

\subsubsection{Structure Factor}\label{sec:structure-factor}

Signs of local structures in the attraction-dominated regime can be
observed in the structure factor at different $\dot\gamma$. The
Structure factor $S(q)$ is defined as:
$S(q) = {N}^{-1}\langle{\rho(q)\rho(-q)}\rangle$, where
$\rho(q)=\sum_{i}^{N}{\rm exp}(i{\bf q}\cdot{\bf r}_i)$ is the Fourier
transform of the number density, N being the total number of
particles.

\begin{figure}[htb]
 \centering
 \includegraphics[width=0.5\textwidth]{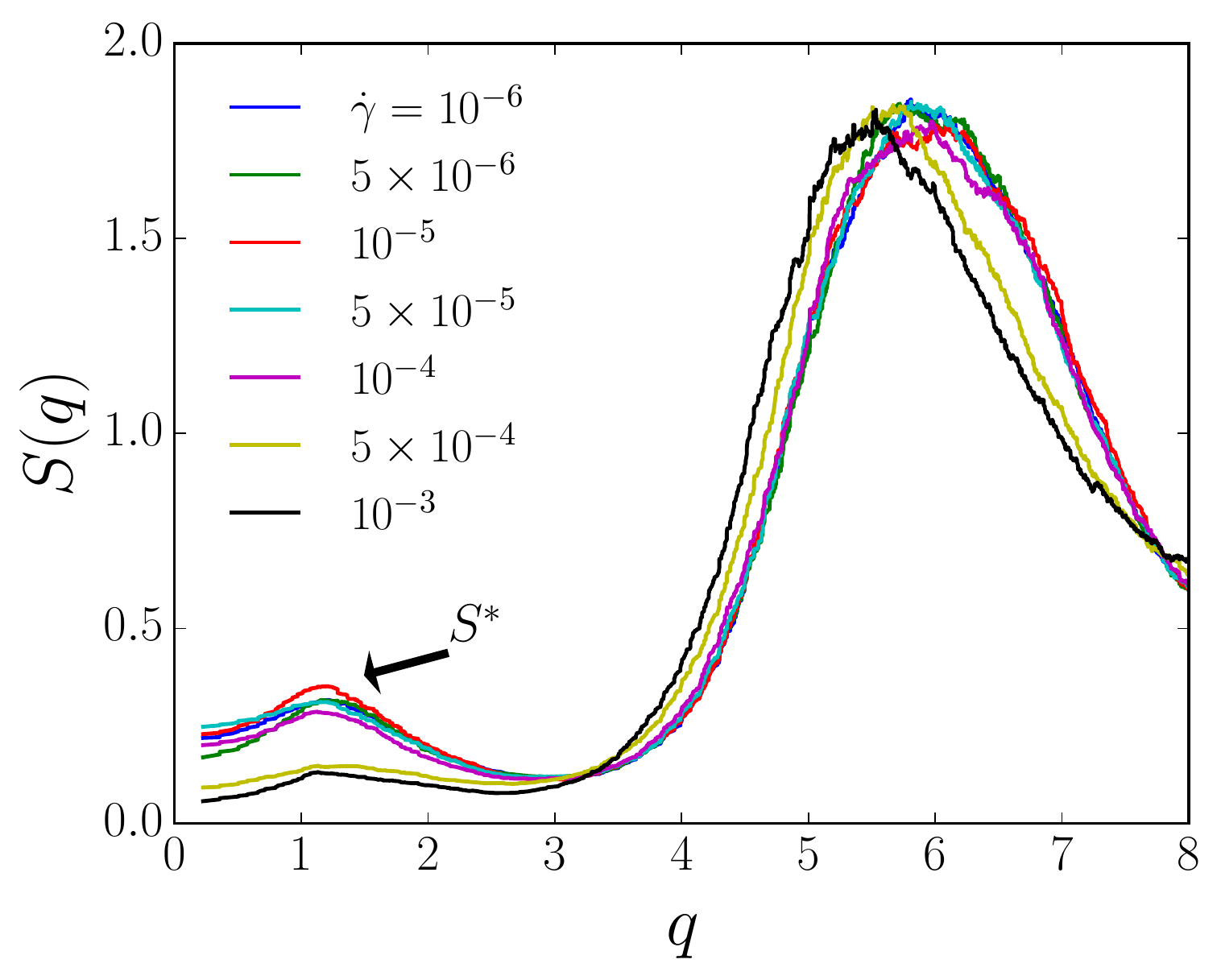}
 \caption{Structure factor $S(q)$ for different $\dot\gamma$ at
   $u=2\times10^{-4}$ and $\phi=0.75$. There is a small peak $S^*$ for
   $q<2$ associated with the local structure and particle clusters in
   the attraction-dominated regime ($\dot\gamma \leq 10^{-4}$). Since data points are noisy
   because of the small system size, they have been smoothed using the
   Savitzky-Golay filter.} \label{fig:sf}
\end{figure}

In Figure \ref{fig:sf}, we plot $S(q)$ at
$\phi=0.75, u=2\times10^{-4}$ for different $\dot\gamma$. One can
observe a small peak at around $q=1$. Such a feature corresponds to
the formation of clusters of particles, induced by shear. Also, we
note that this peak disappears at large shear-rate, implying that this
is due to structures formed by the interplay of small shear and
attractive interactions.

\begin{figure}[htb]
 \centering
 \includegraphics[width=0.45\textwidth]{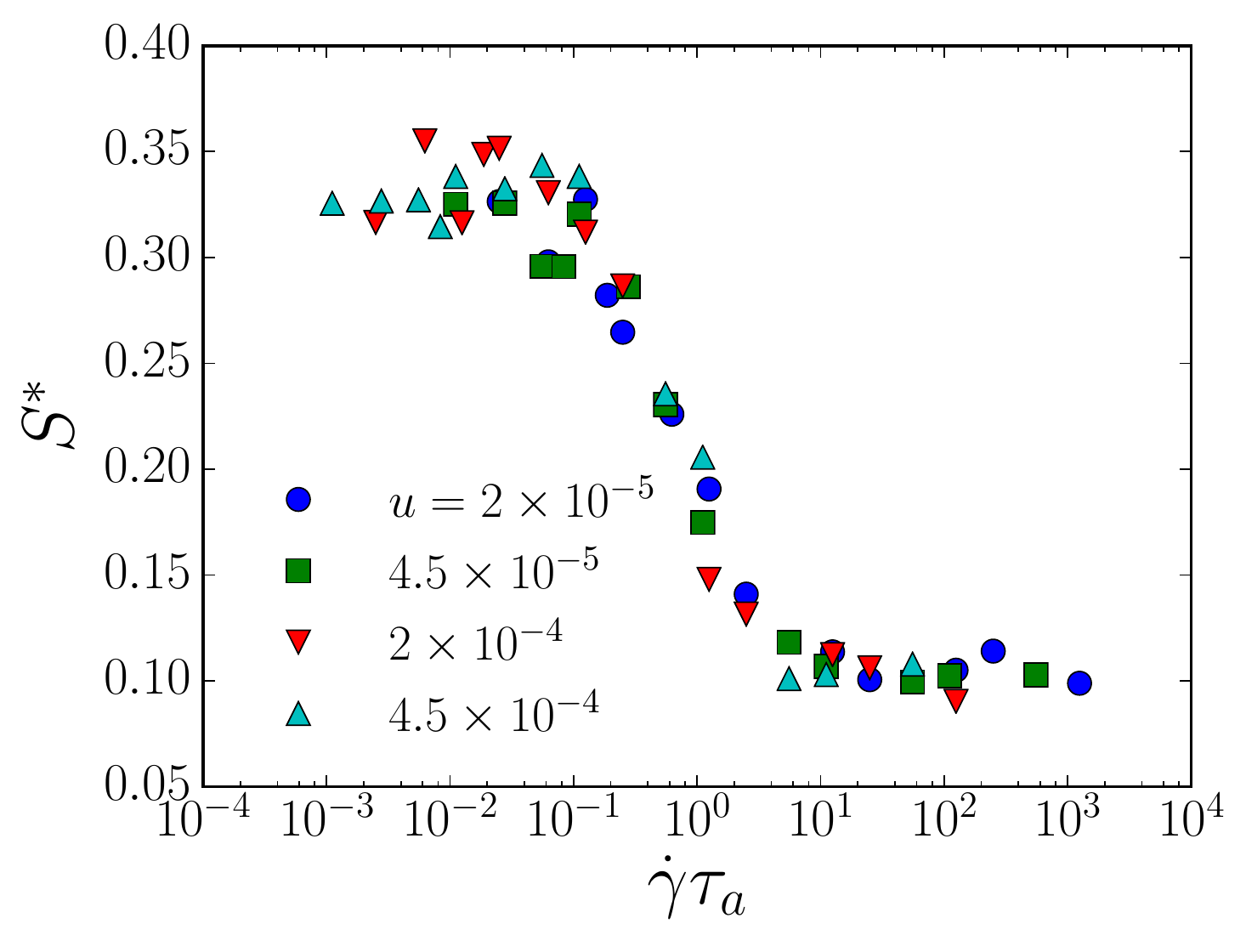}
 \caption{The first maximum in the structure factor, $S^*$ as a
 function of $\dot\gamma \tau_a$ for a system at $\phi=0.75$ and
 different attraction strength. The attractive timescale $\tau_a$
 is used to rescale the strain rate $\dot\gamma$.}
 \label{fig:sf-max}
\end{figure}

With the peak appearing at a certain wavenumber $q^*$, we can define a
peak height via $S^*=S(q^*)$.  In Figure \ref{fig:sf-max}, we show how
$S^*$-data for a range of attraction strengths collapse, if the
shear-rate is rescaled with $\tau_a$. This clearly demonstrates the
distinct effect that attractive forces have in determining the
microstructure over large lengthscales and re-emphasizes the role of
the attractive time-scale $\tau_a$.

\begin{figure}[htb]
 \centering
 \includegraphics[width=0.45\textwidth]{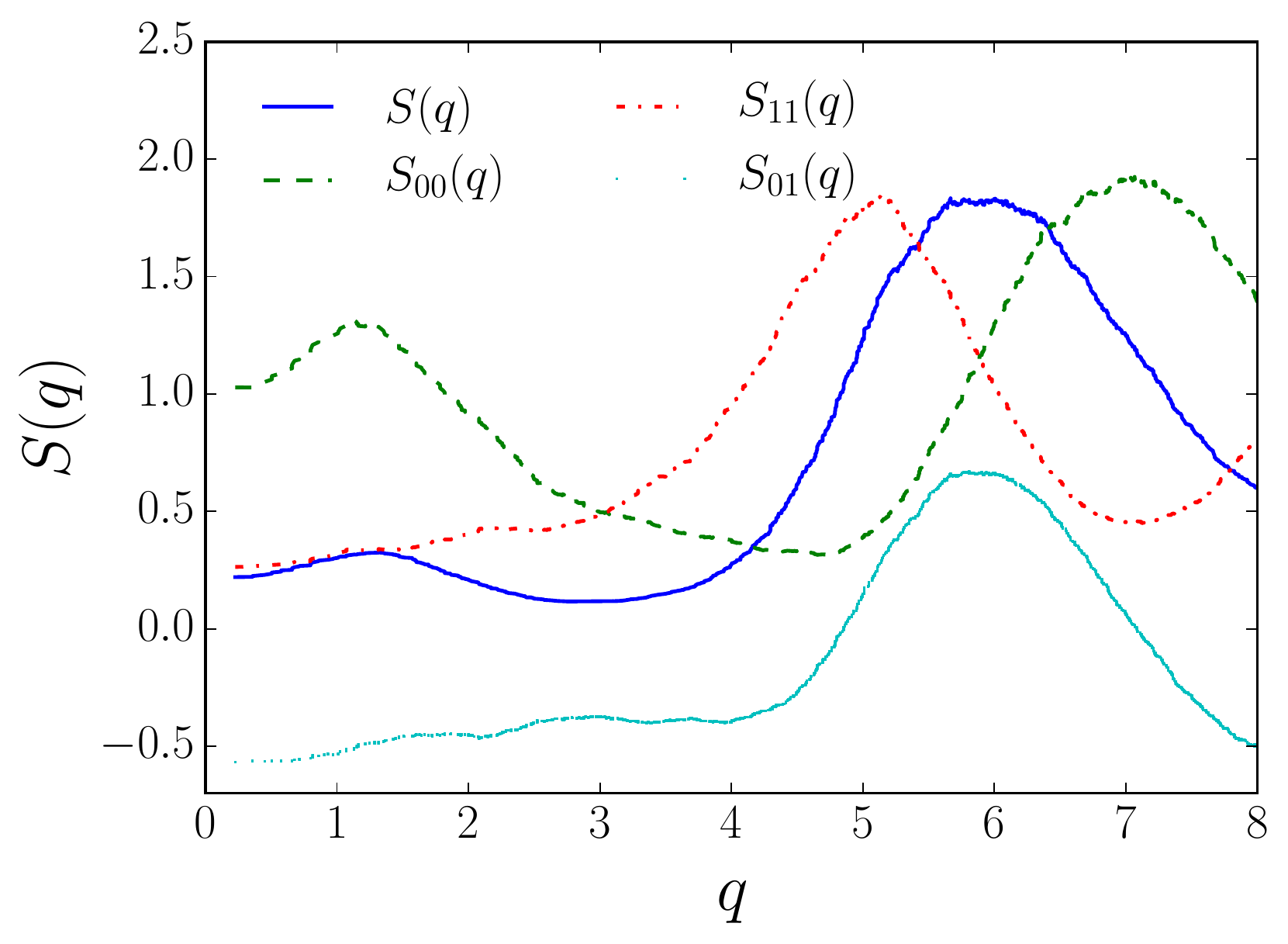}
 \caption{Partial structure factor $S(q)$ for a system at
   $u=2\times10^{-4}$, $\phi=0.75$ and $\dot\gamma=10^{-6}$. $S_{00}$
   corresponds to the smaller particles, while $S_{11}(q)$ measures
   the structure factor of the larger particles.} \label{fig:sf-comp}
\end{figure}

\begin{figure}[htb]
 \centering
 \includegraphics[width=0.45\textwidth]{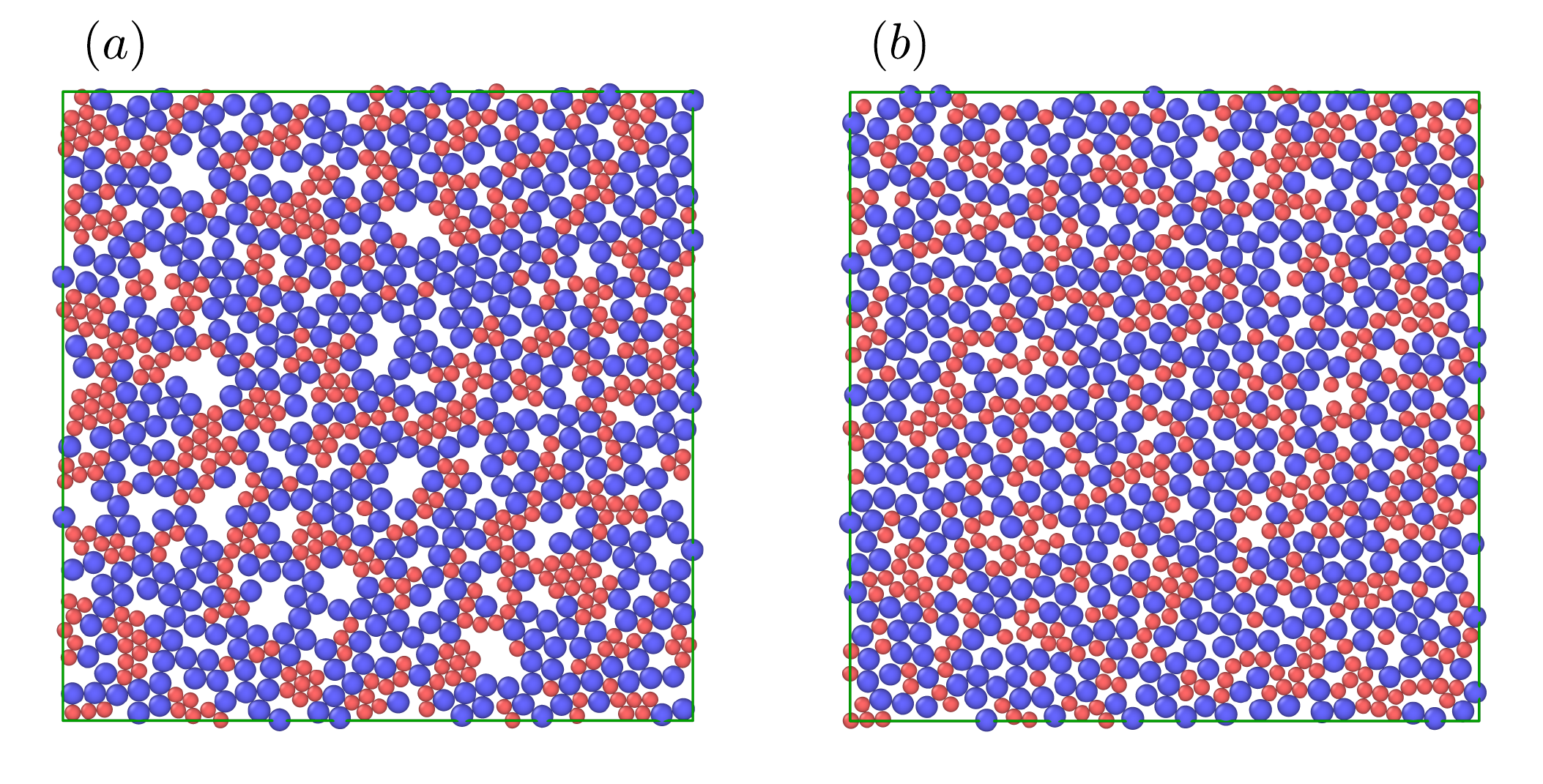}
 \caption{Snapshots of the sheared system for two different
   strainrates ((a) $\dot\gamma=10^{-6}$ and (b) $\dot\gamma=10^{-3}$)
   illustrating the low-$q$ peak observed in the structure factor. For
   the smaller strainrate, in the attraction-dominated regime, the
   small particles cluster and the density is rather inhomogeneous, as
   compared to the larger strainrate, in the repulsion-dominated
   regime.} \label{fig:sf-snapshot}
\end{figure}

For a binary mixture, one can further look at the partial structure
factors,
$S^{\gamma\nu}(q)=\langle{\rho^{\gamma}(q)\rho^{\mu}(-q)}\rangle$,
with
$\rho^{\nu}(q)=\sum_{i}^{N^{\nu}}{\rm exp}(i{\bf q}\cdot{\bf r}_i)$
corresponding to the Fourier transforms of the partial density fields.
$N^{\nu}=N/2$ is the number of the sub-population belonging to labels
$\nu=0,1$, corresponding to small and large particles, respectively.
In terms of the partial structure factors, the total structure factor
can be written as:
\begin{equation}
\label{eq:sf-tot-partial}
 S(q) = {1\over{2}}[S^{00}(q)+S^{11}(q)]+S^{01}(q)
\end{equation}
The partial sructure factors are shown in Figure \ref{fig:sf-comp},
which shows that the low-$q$ peak in $S_{00}(q)$ is more pronounced
than in $S_{11}$. This implies that smaller particles contribute
stronger to the low $q$ peak in $S(q)$, i.e. the mesoscale clustering
under shear is caused by the spatial organisation of these small
particles (see Fig.~\ref{fig:sf-snapshot}). It is tempting to
attribute this effect to some sort of phase separation under
shear. However, we did not observe any substantial growth of these
meso-clusters on the time-scales accessible to our simulations.

\subsubsection{Microscopic dynamics}\label{sec:microscopic-dynamics}

\begin{figure}[htb]
 \centering \includegraphics[width=0.5\textwidth]{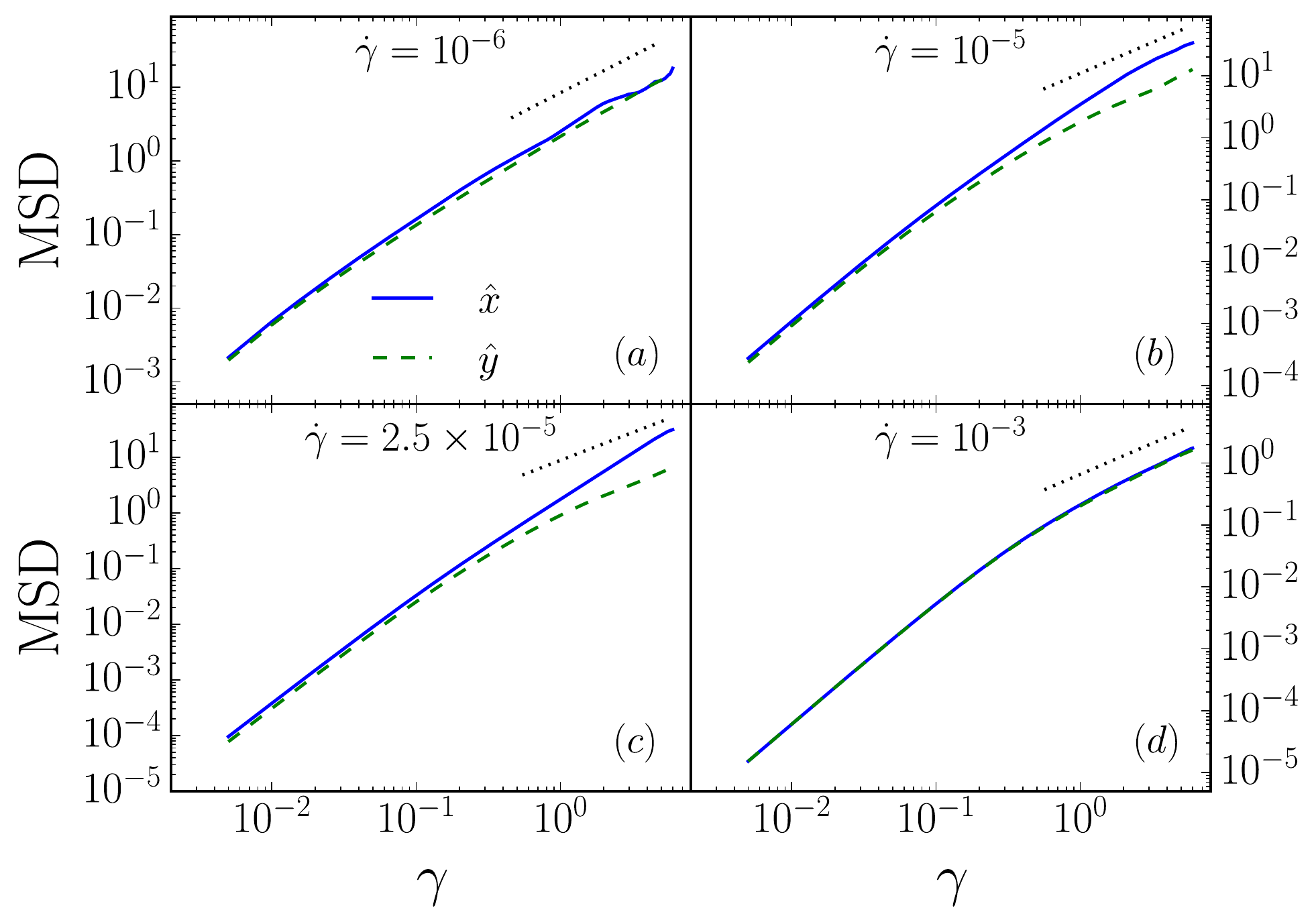}
 \caption{Non-affine mean-squared displacement (MSD) for a system (a)
   close to the yield stress, (b) and (c) around the minimum in the
   flow curve where the attractive and repulsive branches meet and (d)
   in the repulsion-dominated regime, ($u=2\times10^{-5}$ and
   $\phi=0.75$). The dotted line indicates diffusive
   behaviour.} \label{fig:msd}
\end{figure}

Similar to the structure of the particles, their dynamics is also
affected by introducing attraction. To study this, we investigate the
particles' non-affine displacements, which correct for the convective
(affine) part of the particle motion, which is induced by the average
flow field. At any time $t$, the non-affine position of a particle can
be written as:

\begin{equation}
  \mathbf{r}_{\text{naff.}}(t) = \mathbf{r}(t) - \mathbf{\hat{x}} \int_{0}^{t} y(t^\prime)\dot\gamma dt^\prime
\end{equation}

where $\mathbf{r}(t)=x(t)\mathbf{\hat{x}} + y(t)\mathbf{\hat{y}}$ is
the position of the particle at time $t$. The second part corresponds
to the convective contribution, where $\hat{x}$ is the unit vector in
shearing direction and $\hat{y}$ is the unit vector in gradient
direction. Using these non-affine positions, we compute the mean
squared displacement (MSD) of the particles, resolved in $\hat{x}$ and
$\hat{y}$ directions. In Figure \ref{fig:msd}, we show the
corrsponding data for $u=2\times10^{-5}, \phi=0.75$.

At large shear rates (Fig.  \ref{fig:msd}(d)), where repulsion
dominates, there is no difference in MSD between the shearing and the
gradient direction. At large strains, the particles diffusive
isotropically. Similarly, near yielding where attraction dominates,
the dynamics also seems to be not dependent on direction - in this
case, diffusive motion sets in at very large strains, which are
somewhat inaccessible to our simulations [Fig. \ref{fig:msd}(a)].  In
contrast, in the intermediate regime, where the flow curve is
non-monotonic, we observe that the long-time MSD is different in the
flow and the transverse (gradient) directions, i.e. an anisotropy
develops. There is an enhancement of non-affine motions in the flow
direction, which we relate to the nearby instability towards the
formation of shearbands.

\begin{figure}[hbt]
 \centering
  \includegraphics[width=0.5\textwidth]{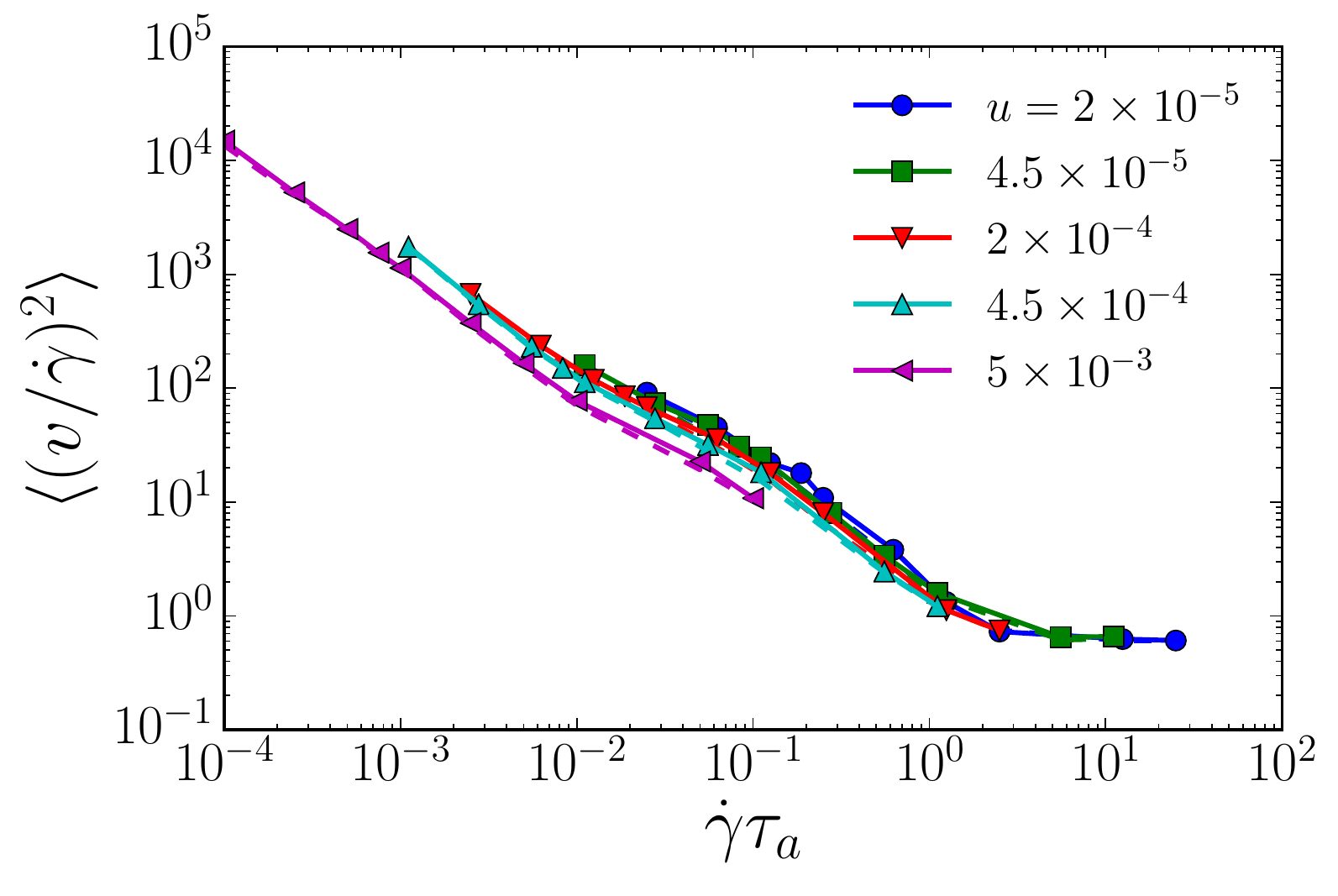}
  \includegraphics[width=0.5\textwidth]{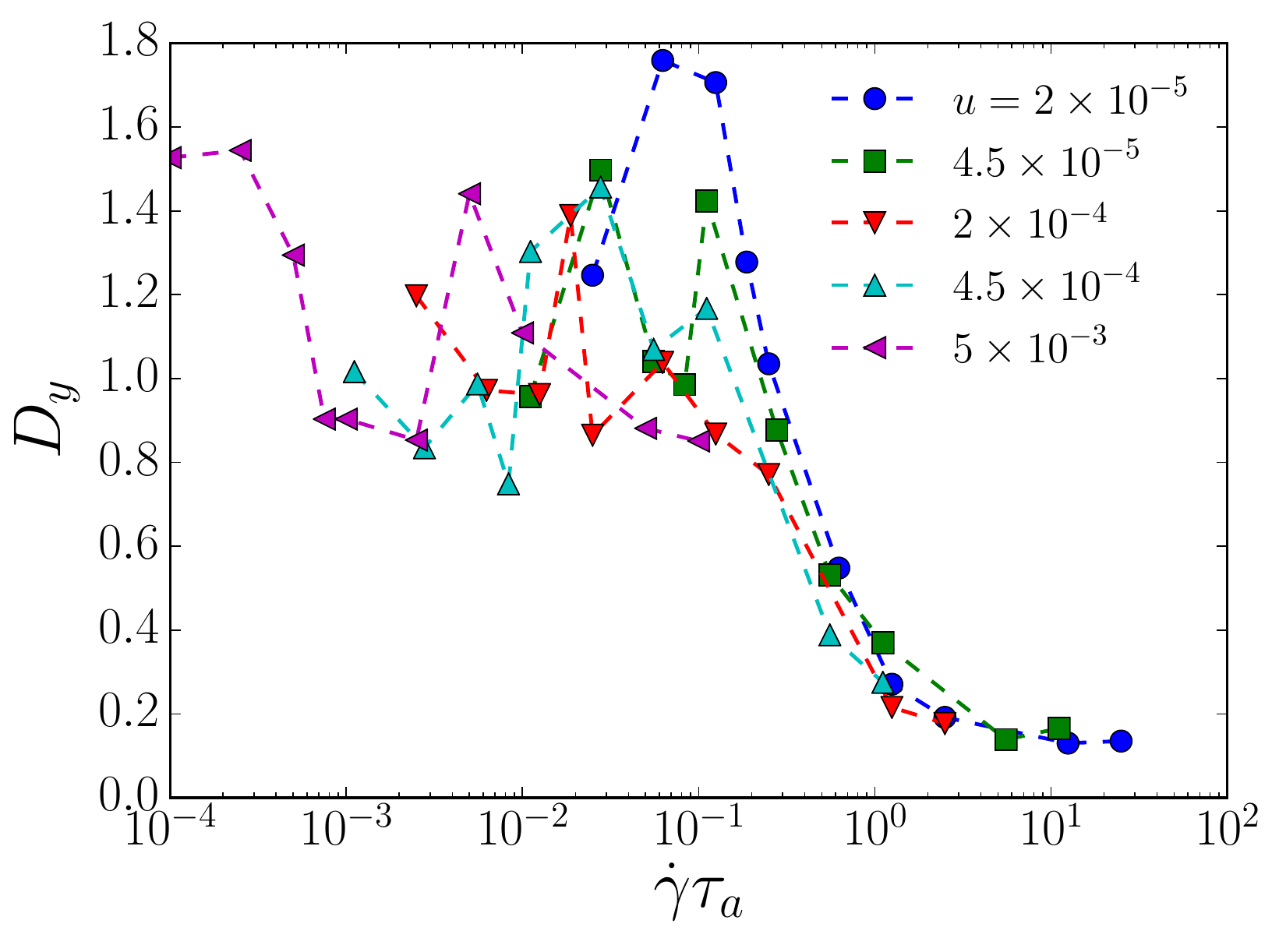}
  \caption{(Top) Variation of non-affine mean-squared velocity with strain rate (scaled by the attractive time-scale) in shearing and gradient directions for a system at
    $\phi=0.75$, for different attraction strengths $u$. While solid
    lines represent the shearing direction, dashed lines correspond to
    the gradient direction. (Bottom) The diffusion constant, measured
    in gradient direction, as a function of scaled shear-rate, for
    different $u$.}
 \label{fig:D}
\end{figure}

To further investigate the nature of the dynamics under shear, we also
measured the mean squared non-affine velocity (${v^2}$) in both directions
as functions of strain rate. The corresponding data is shown in the top
panel of  Figure \ref{fig:D}. In order to compare the mean squared non-affine velocity
of systems at different strain rates, we consider the rescaled
quantity $({v}/\dot\gamma)^2$. We observe that this quantity,
which measures the extent of the particles' non-affine motion over
very short time, increases rapidly with decreasing shear-rate.
We also measure the diffusion coefficient ($D_y$) in the gradient
direction, which is shown in the bottom panel of   Figure \ref{fig:D}.
This quantifies the extent of non-affine motion over
long time and similar to  $({v}/\dot\gamma)^2$,
has a higher value at smaller shear-rates. The large
scatter in the data reflects the approximate nature of calculating the
diffusion constant in a regime, where real diffusion is hardly
reached. Thus, these observables demonstrate that in the attraction
dominated regime, non-affine motion dominates over affine motion,
implying that the latter is more energetically costly.

\subsection{Characterizing shear bands}\label{sec:char-shear-bands}

The non-monotonic shape of $\sigma(\dot\gamma)$ is a signature of a
mechanical instability
\cite{SchallShearBand2010,FieldingRepProgPhys2014,CoussotShearBand2009,PicardPRE2002}.
However, no shear bands are observed in a system of size $N=1000$ for
which the flow curves were presented in Figure \ref{fig:fc-diffphi}.
It is known that for shear localization to occur, the wavelengths of
the unstable modes should be smaller than the system size
\cite{DhontPRE99}. The localization of flow into a shear band is
demonstrated in Figure \ref{fig:snap-velf-z}, which illustrates the
velocity field (and corresponding connectivity profile) for a system
of size $N=2\times10^4$ at suitable state parameters, viz.
$u=2\times10^{-5}$, $\phi=0.82$ and $\dot\gamma=2.5\times10^{-6}$.

\begin{figure}[htb]
	\includegraphics[width=0.5\textwidth]{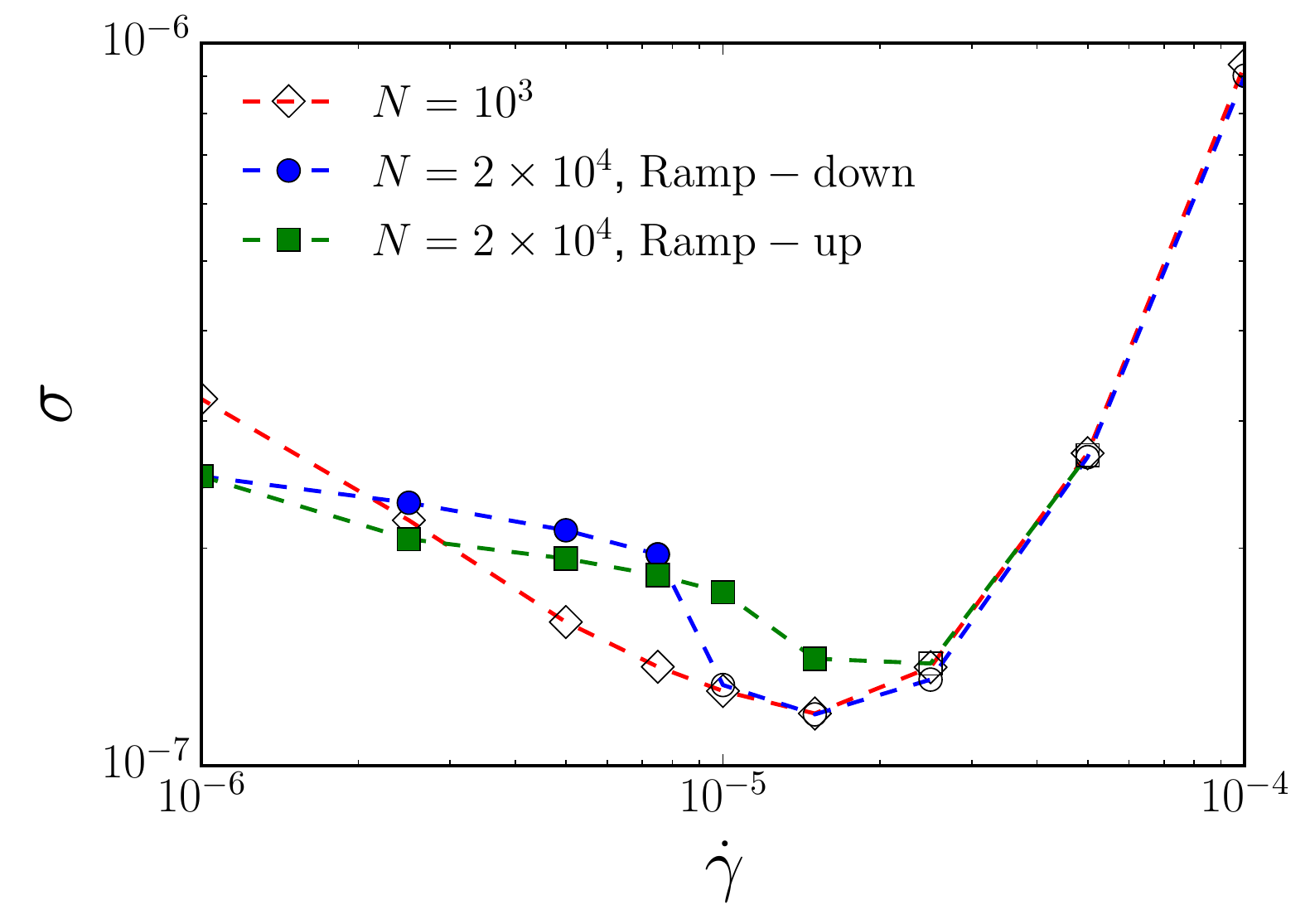}
	\caption{Flow curves ($\hat\sigma=\sigma d/\epsilon u$) for
          different system sizes $N$, at $\phi = 0.82$ and
          $u=2\times10^{-5}$. The non-monotonic part of the flow curve
          gets smaller as system size increases and the system is
          sheared longer. Filled symbols indicate a shear banded flow,
          open symbols correspond to homogeneous flow.}
	\label{fig:fcDiffN}
\end{figure}

We therefore compare how the flow response compares across different
system sizes. In Figure \ref{fig:fcDiffN}, we show
$\sigma(\dot\gamma)$ for two different system sizes, viz.
$N=1000, 20000$. For $N=20000$, we start shearing the system with a
random initial configuration at $\dot\gamma=10^{-4}$, ramp it down
until $\dot\gamma=10^{-6}$ (data shown in circles) and then again ramp
it up (data shown in squares). At each $\dot\gamma$, the system is
sheared for $\Delta\gamma=20$ (except at $\dot\gamma=10^{-6}$, where
we choose $\Delta\gamma=12$).  Such large strain windows at each
$\dot\gamma$ try to ensure that we obtain a steady state response at
each state point. The obtained $\sigma(\dot\gamma)$ data, for the
ramp-down and ramp-up, is shown in Figure \ref{fig:fcDiffN}. We
observe hysteretic effects similar to findings in recent experiments
using the same protocol \cite{DivouxPRL2013}. Further, we mark with
filled symbols the state points at which shear-banding is observed. In
comparison to the data for $N=1000$, the flow curves deviate in the
small $\dot\gamma$ regime.  When the applied shear-rate is ramped
down, the non-monotonicity is {more} pronounced in the regime of small
$\dot\gamma$. And, when the applied shear-rate is ramped up, the
non-monotonity is nearly suppressed.  Further, shear-localization is
also visible over a larger range of shear-rates during the ramp-up
regime; one needs to go to larger shear-rates to fluidize the
solid. It has been suggested that, in the thermodynamic limit, the
stress-decreasing part of the flow curve will be replaced by a
straight line \cite{DhontPRE99, CoussotShearBand2009}; our
observations are consistent with that trend.

\begin{figure}[htb]
	\includegraphics[width=0.45\textwidth]{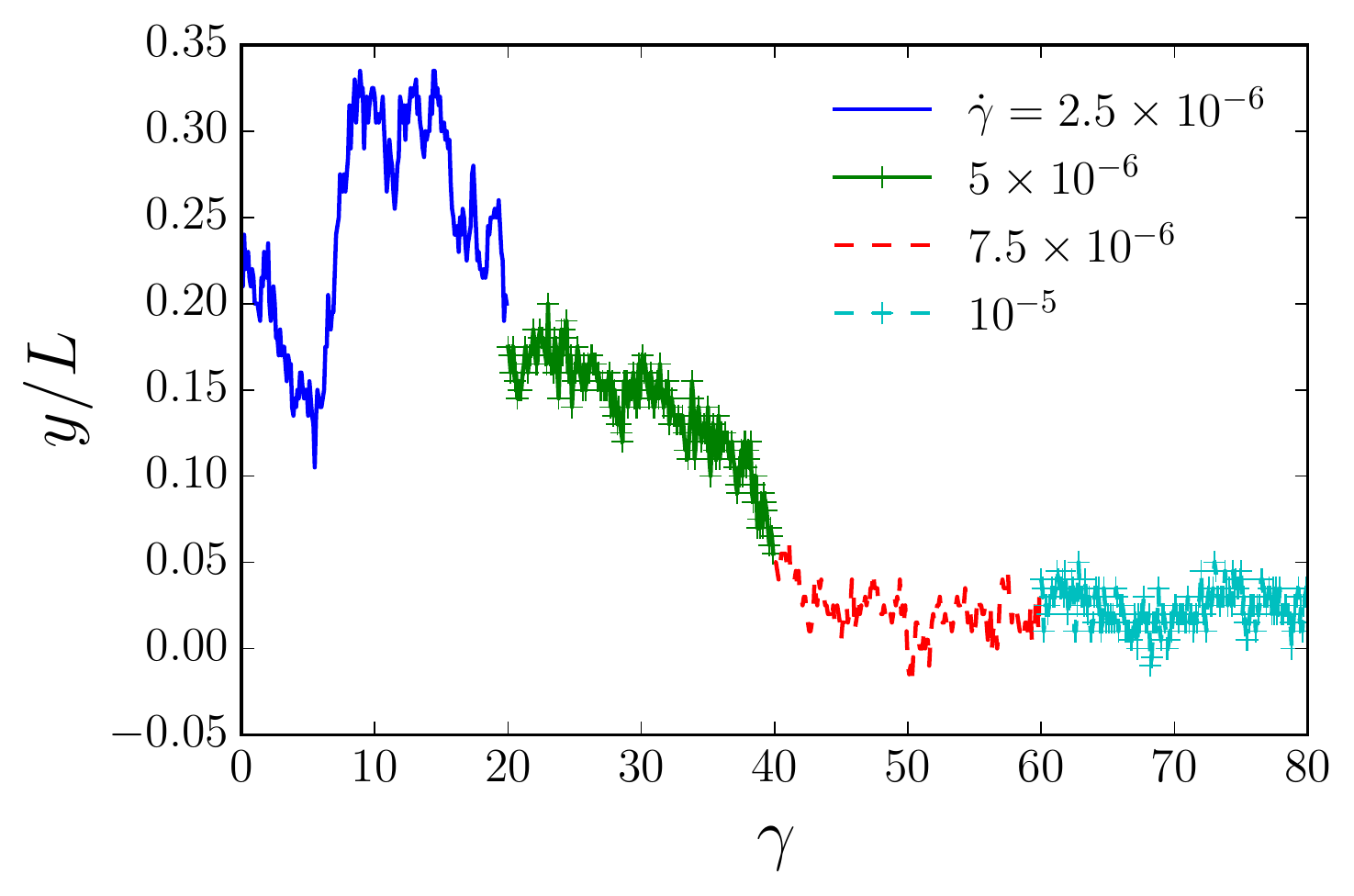}
	\caption{Position of the center of shear band in time for ramp-up simulations.}
	\label{fig:bandmotion}
\end{figure}

We also observe how the spatial location of the fluidized band, in the
direction transverse to the applied shear, changes as a function of
time, by monitoring the position of the centre of the band. In Figure
\ref{fig:bandmotion}, we show how the location varies as the applied
shear-rate is ramped up. At small applied $\dot\gamma$, the position
of the band fluctuates, although the $\Delta\gamma$ is not large
enough for the band to traverse the entire system. As $\dot\gamma$ is
ramped up, there is a contrast in mobility with the band becoming
nearly static at $\dot\gamma^\star$ where the flow curve has a minimum.
This is consistent with our results for the non-affine motion of
particles in small systems (see Fig. \ref{fig:D}) and can explain the
existence of large history effects in the ramping simulations. Only on
timescales large enough for the fluid band to traverse the entire
system, can structures in the solid band be erased. On smaller
time-scales these textures remain and reflect the properties of the
system at the previous strain rate probed.

\begin{figure}[htb]
	\includegraphics[width=0.5\textwidth]{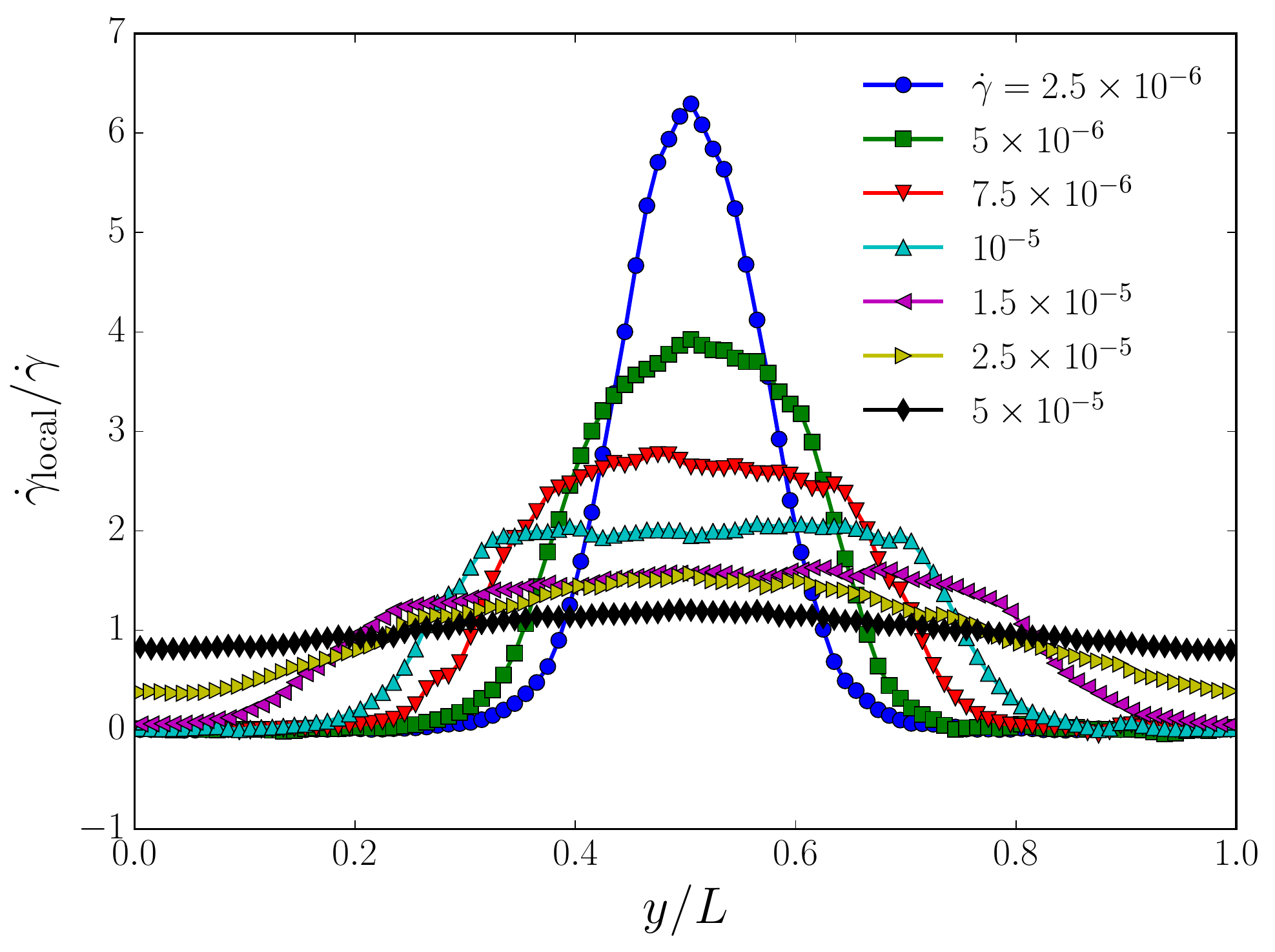}
	\includegraphics[width=0.5\textwidth]{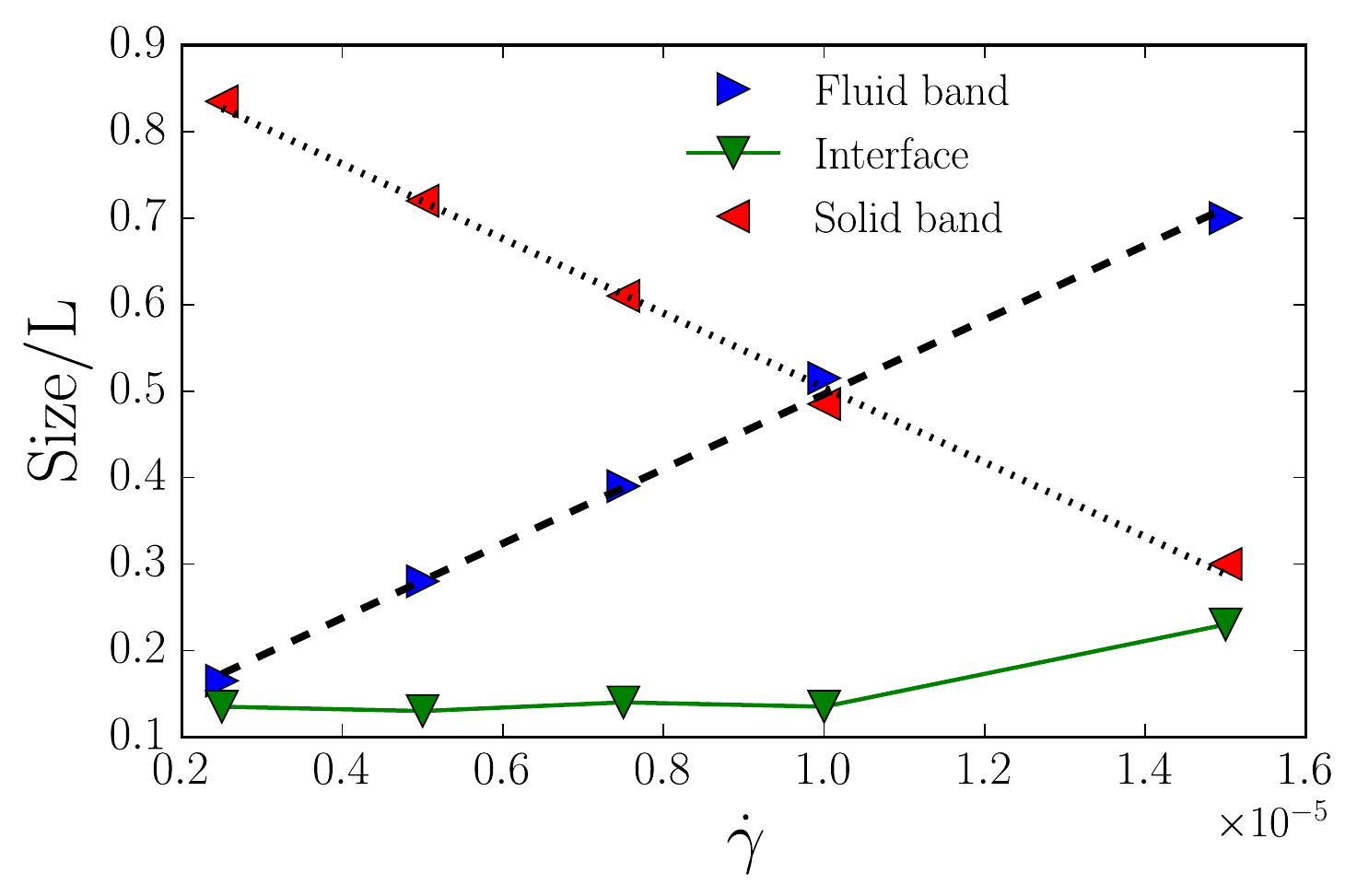}
	\caption{(Top), centralized shear bands for different
          strain rates in ramp-up branch. (Bottom), shear bands
          and the interface width in units of system size, as a
          function of imposed strain rate $\dot\gamma$. The dashed
          line is the fitted linear function to the size of fluid
          band,
          $S_f(\dot\gamma)=4.31\times10^{4}\dot\gamma+6.51\times10^{-2}$. The
          dotted line presents the fitted function to the size of
          solid band which is by definition
          $S_s(\dot\gamma)=1-S_f(\dot\gamma)$.}
	\label{fig:shearbands}
\end{figure}

Finally, we explore how the spatial profile of the shear-bands change,
when the external strain rate is varied. In the top panel of Figure
\ref{fig:shearbands}, we show the spatial profile of local shear-rates
($\dot\gamma_{\rm local}$), normalised by the imposed $\dot\gamma$.
At the smallest shear-rate, we have a very localised band of large
fluidization.  With increasing shear-rate, we observe that the height
of this spatial profile decreases, implying that the fluidized region
has less contrast in flow-rate with the solid-like region. Also, the
width of this region broadens and the shear-band finally disappears in
the regime where repulsion dominates.  In all these cases, we have
checked and found that the stress generated in the system is spatially
uniform, albeit with minor fluctuations.

{F}rom the spatial profiles of the shear-bands, we compute the width of
the solid-like and fluid-like regions, as well as the interface of the
shear-band. This data is shown in the bottom panel of Figure
\ref{fig:shearbands}.  As described above, the width of the solid-like
region decreases linearly and, similarly, the liquid-like region
increases with increasing $\dot\gamma$. On the contrary, the width of
the interface remains nearly constant, consistent with earlier
numerical observations \cite{MartensSoftMatter2012}. Thus it might
actually reflect an intrinsic material property, e.g. a shear
curvature viscosity \cite{DhontPRE99}.

\section{Conclusion}\label{sec:conclusion}

We have studied the rheological response of an athermal system of
particles, having weak attractive interactions, by scanning across a
wide range of packing fraction ($\phi$), attraction strength ($u$) and
imposed shear-rates ($\dot{\gamma}$). These extensive simulations
reveal that at vanishing shear-rates and weak attractions, a fragile
isostatic solid exists even if we go to $\phi \ll \phi_J$. Further,
with increasing shear-rates, even at these low $\phi$, non-monotonic
flow curves occur which (in large enough systems) lead to persistent
shear-bands. Our exploration of parameters allows us to draw up the
regime in $u-\phi$ where such shearbanding is to be expected, which we
observe to be spanning a large parameter window. The non-monotonic
flow curves are also associated with a non-monotonic dependence of the
macroscopic friction, $\mu$, on imposed shear-rate, with the maximum
in the resistance to shear occurring at the exact point where the
minimum in the flow curves occur. The low shear-rate regime, where
attractive interactions dominate the rheological response, is also
characterized by enhanced non-affine dynamics. Finally, we demonstrate
that the non-monotonicity in the flow curves is enhanced if one tunes
the dissipation timescale to probe the under-damped regime of the
dynamics.

The solid-like response of any material is characterized by the
yielding stress threshold, $\sigma_y$. Similarly, we define a
potential energy threshold, $E_y$. Just like for $\sigma_y$, we
demonstrate the existence of scaling relationships for $E_y$. {Where}
$\phi > \phi_J$, $E_y \propto \delta{\phi}^\beta$, with $\beta \approx
2.1$. Also, for $\phi < \phi_J$, we show that $|E_y| \sim u^2$,
i.e. it is determined by the strength of the attractive interaction
between the particles.

The macroscopic rheological response of such materials is further
probed at the particle level by measuring the structure factor,
$S(q)$.  It exhibits a peak at small $q$, implying clustering of
particles, at small shear-rates.  By studying the partial structure
factor, we conclude that the mesoscale clustering induced by
shear consists primarily of the smaller particles.


Although non-monotonic flow curves are obtained for $N=1000$, we need
to explore larger system sizes, e.g. $N=20000$, in order for
shear-bands to form.  This is due to the fact that sufficiently large
systems are necessary for accommodating the spatially heterogeneous
flow. Precursors of this banding transition can, however, be observed
in a pronounced anisotropy of the particle dynamics, with the
deviatoric velocity as well as the diffusion constant in flow
direction enhanced as compared to the gradient direction.  We also
observe that for such large systems, the flow curves deviate from
those obtained in smaller systems. Further, similar to recent
experiments, hysteretic effects are observed when the applied
shear-rate is ramped down and then ramped up. We explain these
features with a small mobility of the fluid band leading to
prohibitively long time-scales necessary to fully erase any memory
present in the solid band. During such a ramping protocol,
shear-banding becomes prominent when the strain rate is ramped up,
with larger shear-rates necessary to fluidize the solid.

In future, extensive experimental studies are necessary for exploring
and validating the observations made via our numerical investigations
regarding the rheological response of weakly attractive particulate
systems.  All these studies have focused on athermal suspensions.
Further explorations are necessary to check how the flow behaviour
changes if one considers Brownian suspensions of such particles,
i.e. whether thermal fluctuations modify the response. Studies are
also necessary to understand the rheology of such suspensions in the
context of more complex flow protocols, which are common in industrial
applications.

\section{Acknowledgments}

We acknowledge financial support by the DFG via the Emmy Noether
Program (No. He 7429/1-1). We also thank Jean-Louis Barrat for useful
discussions.

\bibliographystyle{unsrt}
\bibliography{e2}

\begin{thebibliography}{10}

\bibitem{vanHeckeJPCM2010}
M~van Hecke.
\newblock {Jamming of soft particles: geometry, mechanics, scaling and
  isostaticity}.
\newblock {\em Journal of Physics: Condensed Matter}, 22(3):033101, 2010.

\bibitem{LiuNagelAnnRevCondMattPhys2010}
Andrea~J. Liu and Sidney~R. Nagel.
\newblock The jamming transition and the marginally jammed solid.
\newblock {\em Annual Review of Condensed Matter Physics}, 1(1):347--369, 2010.

\bibitem{ClausPinakiSoftmatter2010}
Claus Heussinger, Pinaki Chaudhuri, and Jean-Louis Barrat.
\newblock Fluctuations and correlations during the shear flow of elastic
  particles near the jamming transition.
\newblock {\em Soft Matter}, 6:3050--3058, 2010.

\bibitem{GrobPRE2014}
Matthias Grob, Claus Heussinger, and Annette Zippelius.
\newblock Jamming of frictional particles: A nonequilibrium first-order phase
  transition.
\newblock {\em Phys. Rev. E}, 89:050201, May 2014.

\bibitem{daCruzPRE2005}
Fr\'ed\'eric da~Cruz, Sacha Emam, Micha\"el Prochnow, Jean-No\"el Roux, and
  Fran\ifmmode \mbox{\c{c}}\else~\c{c}\fi{}ois Chevoir.
\newblock Rheophysics of dense granular materials: Discrete simulation of plane
  shear flows.
\newblock {\em Phys. Rev. E}, 72:021309, Aug 2005.

\bibitem{IraniPRL2014}
Ehsan Irani, Pinaki Chaudhuri, and Claus Heussinger.
\newblock Impact of attractive interactions on the rheology of dense athermal
  particles.
\newblock {\em Phys. Rev. Lett.}, 112:188303, May 2014.

\bibitem{PinakiPRE2012}
Pinaki Chaudhuri, Ludovic Berthier, and Lyd\'{e}ric Bocquet.
\newblock Inhomogeneous shear flows in soft jammed materials with tunable
  attractive forces.
\newblock {\em Phys. Rev. E}, 85:021503, Feb 2012.

\bibitem{SundaresanPRE2014}
Yile Gu, Sebastian Chialvo, and Sankaran Sundaresan.
\newblock Rheology of cohesive granular materials across multiple dense-flow
  regimes.
\newblock {\em Phys. Rev. E}, 90:032206, Sep 2014.

\bibitem{RahbariPRE2010}
S.~H.~Ebrahimnazhad Rahbari, J.~Vollmer, S.~Herminghaus, and M.~Brinkmann.
\newblock Fluidization of wet granulates under shear.
\newblock {\em Phys. Rev. E}, 82:061305, Dec 2010.

\bibitem{SinghPRE2014}
Abhinendra Singh, Vanessa Magnanimo, Kuniyasu Saitoh, and Stefan Luding.
\newblock Effect of cohesion on shear banding in quasistatic granular
  materials.
\newblock {\em Phys. Rev. E}, 90:022202, Aug 2014.

\bibitem{HornbakerNature1997}
{Hornbaker D. J.}, {Albert R.}, {Albert I.}, {Barabasi A.-L.}, and {Schiffer
  P.}
\newblock {What keeps sandcastles standing?}
\newblock {\em Nature}, 387(6635):765--765, June 1997.

\bibitem{HerminghausAdvPhys2005}
S.~Herminghaus.
\newblock Dynamics of wet granular matter.
\newblock {\em Advances in Physics}, 54(3):221--261, 2005.

\bibitem{MitaraiAdvPhys2006}
Namiko Mitarai and Franco Nori.
\newblock Wet granular materials.
\newblock {\em Advances in Physics}, 55(1-2):1--45, 2006.

\bibitem{CastellanosAdvPhys2005}
A.~Castellanos.
\newblock The relationship between attractive interparticle forces and bulk
  behaviour in dry and uncharged fine powders.
\newblock {\em Advances in Physics}, 54(4):263--376, 2005.

\bibitem{RoyerNature2009}
{Royer John R.}, {Evans Daniel J.}, {Oyarte Loreto}, {Guo Qiti}, {Kapit Eliot},
  {Mobius Matthias E.}, {Waitukaitis Scott R.}, and {Jaeger Heinrich M.}
\newblock {High-speed tracking of rupture and clustering in freely falling
  granular streams}.
\newblock {\em Nature}, 459(7250):1110--1113, June 2009.
\newblock 10.1038/nature08115.

\bibitem{CoussotSoftMatter2007}
P.~Coussot.
\newblock Rheophysics of pastes: a review of microscopic modelling approaches.
\newblock {\em Soft Matter}, 3:528--540, 2007.

\bibitem{Moller5139}
Peder Moller, Abdoulaye Fall, Vijayakumar Chikkadi, Didi Derks, and Daniel
  Bonn.
\newblock An attempt to categorize yield stress fluid behaviour.
\newblock {\em Philosophical Transactions of the Royal Society of London A:
  Mathematical, Physical and Engineering Sciences}, 367(1909):5139--5155, 2009.

\bibitem{GreggOHernPRL2008}
Gregg Lois, Jerzy Blawzdziewicz, and Corey~S. O'Hern.
\newblock Jamming transition and new percolation universality classes in
  particulate systems with attraction.
\newblock {\em Phys. Rev. Lett.}, 100:028001, Jan 2008.

\bibitem{SchallShearBand2010}
Peter Schall and Martin van Hecke.
\newblock {Shear Bands in Matter with Granularity}.
\newblock {\em Annual Review of Fluid Mechanics}, 42(1):67--88, 2010.

\bibitem{CoussotShearBand2009}
G.~Ovarlez, S.~Rodts, X.~Chateau, and P.~Coussot.
\newblock {Phenomenology and physical origin of shear localization and shear
  banding in complex fluids}.
\newblock {\em Rheologica Acta}, 48(8):831--844, 2009.

\bibitem{BecuPRL2006}
Lydiane B\'ecu, S\'ebastien Manneville, and Annie Colin.
\newblock Yielding and flow in adhesive and nonadhesive concentrated emulsions.
\newblock {\em Phys. Rev. Lett.}, 96:138302, Apr 2006.

\bibitem{OvarlezEPL2010}
G.~Ovarlez, K.~Krishan, and S.~Cohen-Addad.
\newblock Investigation of shear banding in three-dimensional foams.
\newblock {\em EPL (Europhysics Letters)}, 91(6):68005, 2010.

\bibitem{DivouxPRL2010}
Thibaut Divoux, David Tamarii, Catherine Barentin, and S\'ebastien Manneville.
\newblock Transient shear banding in a simple yield stress fluid.
\newblock {\em Phys. Rev. Lett.}, 104:208301, May 2010.

\bibitem{FieldingRepProgPhys2014}
S~M Fielding.
\newblock Shear banding in soft glassy materials.
\newblock {\em Reports on Progress in Physics}, 77(10):102601, 2014.

\bibitem{KadauPT2003}
D.~Kadau, G.~Bartels, L.~Brendel, and D.E. Wolf.
\newblock Pore stabilization in cohesive granular systems.
\newblock {\em Phase Transitions}, 76(4-5):315--331, 2003.

\bibitem{GilabertPRE2007}
F.~A. Gilabert, J.-N. Roux, and A.~Castellanos.
\newblock Computer simulation of model cohesive powders: Influence of
  assembling procedure and contact laws on low consolidation states.
\newblock {\em Phys. Rev. E}, 75:011303, Jan 2007.

\bibitem{KhamsehPRE2015}
Saeed Khamseh, Jean-No\"el Roux, and Fran\ifmmode
  \mbox{\c{c}}\else~\c{c}\fi{}ois Chevoir.
\newblock Flow of wet granular materials: A numerical study.
\newblock {\em Phys. Rev. E}, 92:022201, Aug 2015.

\bibitem{ClausPRE2013}
Claus Heussinger.
\newblock Shear thickening in granular suspensions: Interparticle friction and
  dynamically correlated clusters.
\newblock {\em Phys. Rev. E}, 88:050201, Nov 2013.

\bibitem{LAMMPS}
http://lammps.sandia.gov/index.html.

\bibitem{DhontPRE99}
Jan K.~G. Dhont.
\newblock A constitutive relation describing the shear-banding transition.
\newblock {\em Phys. Rev. E}, 60:4534--4544, Oct 1999.

\bibitem{PicardPRE2002}
Guillemette Picard, Armand Ajdari, Lyd\'{e}ric Bocquet, and Fran\ifmmode
  \mbox\c{c}\else~\c{c}\fi{}ois Lequeux.
\newblock Simple model for heterogeneous flows of yield stress fluids.
\newblock {\em Phys. Rev. E}, 66:051501, Nov 2002.

\bibitem{TeitelPRL2014}
Daniel V{\aa}gberg, Peter Olsson, and S.~Teitel.
\newblock Universality of jamming criticality in overdamped shear-driven
  frictionless disks.
\newblock {\em Phys. Rev. Lett.}, 113:148002, Oct 2014.

\bibitem{OHernPRE2003}
Corey~S. O'Hern, Leonardo~E. Silbert, Andrea~J. Liu, and Sidney~R. Nagel.
\newblock Jamming at zero temperature and zero applied stress: The epitome of
  disorder.
\newblock {\em Phys. Rev. E}, 68:011306, Jul 2003.

\bibitem{ClausPRL2007}
O.~Lieleg, M.~M. A.~E. Claessens, C.~Heussinger, E.~Frey, and A.~R. Bausch.
\newblock Mechanics of bundled semiflexible polymer networks.
\newblock {\em Phys. Rev. Lett.}, 99:088102, Aug 2007.

\bibitem{WyartPRL2008}
M.~Wyart, H.~Liang, A.~Kabla, and L.~Mahadevan.
\newblock Elasticity of floppy and stiff random networks.
\newblock {\em Phys. Rev. Lett.}, 101:215501, Nov 2008.

\bibitem{RognonRouxFLM2008}
P.~Rognon, J.~Roux, M.~Naa\"im, and F~Chevoir.
\newblock Dense flows of cohesive granular materials.
\newblock {\em Journal of Fluid Mechanics}, 596:21--47, 1 2008.

\bibitem{BergerEPL2015}
Nicolas Berger, Emilien Az{\'e}ma, Jean-Fran{\c c}ois Douce, and Farhang
  Radjai.
\newblock Scaling behaviour of cohesive granular flows.
\newblock {\em EPL (Europhysics Letters)}, 112(6):64004, 2015.

\bibitem{SalernoPRE2013}
K.~Michael Salerno and Mark~0. Robbins.
\newblock Effect of inertia on sheared disordered solids: Critical scaling of
  avalanches in two and three dimensions.
\newblock {\em Phys. Rev. E}, 88:062206, December 2013.

\bibitem{NicolasPRL2016}
Alexandre Nicolas, Jean-Louis Barrat, and J\"org Rottler.
\newblock Effects of inertia on the steady-shear rheology of disordered solids.
\newblock {\em Phys. Rev. Lett}, 116:058303, February 2016.

\bibitem{DivouxPRL2013}
Thibaut Divoux, Vincent Grenard, and S\'ebastien Manneville.
\newblock Rheological hysteresis in soft glassy materials.
\newblock {\em Phys. Rev. Lett.}, 110:018304, January 2013.

\bibitem{MartensSoftMatter2012}
Kirsten Martens, Lyderic Bocquet, and Jean-Louis Barrat.
\newblock {Spontaneous formation of permanent shear bands in a mesoscopic model
  of flowing disordered matter}.
\newblock {\em Soft Matter}, 8:4197--4205, 2012.

\end{thebibliography}
\end{document}